\documentclass[aps,twocolumn,a4paper,superscriptaddress]{revtex4-1}
\usepackage{tensor}
\usepackage{graphicx}
\usepackage{amsmath}
\usepackage{amssymb}
\usepackage{enumerate}
\usepackage{subfigure}
\usepackage{tabularx}
\usepackage[colorlinks=true, pdfstartview=FitV, linkcolor=blue, citecolor=red, urlcolor=black, breaklinks=true]{hyperref}
\usepackage{amsthm}
\usepackage{epstopdf}
\usepackage{grffile}
\usepackage{lipsum}
\usepackage{float}
\usepackage{gensymb}
\usepackage{footmisc}
\usepackage[utf8]{inputenc}
\usepackage{comment}

\theoremstyle{definition}

\newcommand{\be}{\begin{equation}}
\newcommand{\ee}{\end{equation}}
\newcommand{\ben}{\begin{eqnarray}}
\newcommand{\een}{\end{eqnarray}}
\newcommand{\bes}{\begin{subequations}}
\newcommand{\ees}{\end{subequations}}
\def\bal#1\eal{\begin{align}#1\end{align}}
\newcommand{\vphi}{\varphi}

\newcommand{\LL}{{\mathcal L}}

\newcommand{\A}{\mathcal{A}}
\newcommand{\D}{\mathcal{D}}
\newcommand{\F}{\mathcal{F}}

\begin{document}
\title{Multilayered Vortices}
\author{D. Bazeia}\affiliation{Departamento de F\'\i sica, Universidade Federal da Para\'\i ba, 58051-970 Jo\~ao Pessoa, PB, Brazil}
\author{M. A. Liao}\affiliation{Departamento de F\'\i sica, Universidade Federal da Para\'\i ba, 58051-970 Jo\~ao Pessoa, PB, Brazil}
\author{M. A. Marques}\affiliation{Departamento de F\'\i sica, Universidade Federal da Para\'\i ba, 58051-970 Jo\~ao Pessoa, PB, Brazil}
\author{R. Menezes}\affiliation{Departamento de Ci\^encias Exatas, Universidade Federal da Para\'\i ba, 58297-000 Rio Tinto, PB, Brazil}
\affiliation{Departamento de F\'\i sica, Universidade Federal da Para\'\i ba, 58051-970 Jo\~ao Pessoa, PB, Brazil}
\begin{abstract}
Vortices are localized planar structures that attain topological stability and can be used to describe collective behavior in a diversity of situations of current interest in nonlinear science. In high energy physics, vortices engender integer winding number and appear under the presence of a local Abelian symmetry. In this work we study vortices in a Maxwell-Higgs  model, in which the gauge symmetry is enhanced to accommodate additional symmetries, responsible to generate localized structures to be used to constrain the vortex structure in a given region in the plane. The main aim is to examine how the vortex profile changes when it inhabits a limited region, an issue of current interest to the study of vortices at the nanometric scale.
\end{abstract}
\maketitle

\section{Introduction}

In 1957, Abrikosov unveiled the existence of vortex lattices in superconductors \cite{Ab}. The procedure is based on the Ginzburg–Landau theory \cite{GL} and involves nonrelativistic fields. However, it is also possible to find relativistic field theories that support localized vortex solutions. The first relativistic model was examined by Nielsen and Olesen \cite{NO} in 1973, described by a complex scalar field minimally coupled with a Maxwell gauge field under the action of the local $U(1)$ symmetry. Vortices seem to be everywhere in nonlinear science. They appear in magnetic materials at the nanometric scale \cite{Sci1,Sci2}, in spinor Bose-Einstein condensates \cite{spinor}, in fluids and in many other systems. In superfluid $^3{\rm He}$, in particular, recent works have shown that the presence of superfluid phases are strongly influenced by mesoscopic confinement \cite{Sci3} and by nanoscale channels \cite{prb}, which can greatly alter the phase diagram by stabilizing broken symmetry phases not observed in bulk samples. Vortices also appear in specific arrangements of living systems \cite{vb1,vb2}, in particular in small droplets of dense bacterial suspensions \cite{vb1}, where the influence of global confinement on collective motion was also identified, connected with the competition between radial confinement, self-propulsion and other factors, with the effect of robustly inducing intriguing steady vortex states.

In high energy physics, vortices are topological structures with the magnetic field giving rise to a magnetic flux which is controlled by an integer number $n \in \mathbb{Z}$ of the basic magnetic flux, which the integer also known as vorticity or winding number. Magnetic vortices are vortices described by the magnetization of magnetic materials, and there they are also localized structures topologically protected by the Pontryagin index. Magnetic vortices are half-skyrmions, that is, they have skyrmion number one-half and belong to the class of topological structures known as magnetic skyrmions \cite{Bog,Nature,NRM}, which are protected by the Pontryagin index with integer value, the skyrmion number. Magnetic vortices and magnetic skyrmions have an important interface when one deals with magnetic structures in magnetic materials at the nanometric scale \cite{Sci1,Sci2,NRM}. 

In this work we focus on the relativistic Maxwell-Higgs system, but we enhance the local $U(1)$ symmetry to the case of $U(1)\times G$, with the extra symmetry governed by $G$, which can be the discrete $Z_2$ symmetry or another local $U(1)$ symmetry. Here, however, we innovate using the second symmetry to describe a topological structure which is localized in the plane, and so capable of entrapping the original vortex into a limited region of the plane, modelling the presence of a geometric constriction. We have examined this for kinks in the real line very recently \cite{BLM}, and in the present work we discuss the much harder case of vortices, which requires the presence of two spatial dimensions and the addition of extra degrees of freedom. Issues concerning the enhancement of the local $U(1)$ symmetry were investigated before in \cite{N1} and more recently in Refs. \cite{N2,N3}, for instance, but here we deal with another possibility, focusing on the entrapment of vortices into geometrically constrained regions in the plane. The subject is of current interest, since the study of nanometrically sized vortexlike structures may induce effects due to the appearance of constraints in the material at the nanometric scale \cite{Sci3,prb}. This is the case, for instance, in \cite{N}, where particular geometric junction was used to create skyrmions from domain walls, and also in \cite{S}, where a geometric constriction is of key importance to demonstrate experimentally the current-driven transformation of domain walls into magnetic skyrmions in a magnetic strip, an issue of direct interest to skyrmion-based spintronics. More recently, in \cite{CP19} the authors investigated the pattern formation of geometrically confined skyrmions, and there they show in particular that the disk-shaped geometry directly contributes for the formation of a multilayered magnetic structure.

The magnetic skyrmions and vortices are localized finite energy configurations that appear in magnetic materials and have been studied with a diversity of applications; see, e.g., the recent review \cite{NRM} and references therein. One interesting mechanism responsible for the formation of skyrmions relies on the Dzyaloshinsky-Moriya interaction \cite{D,M,Bog}, which is induced  by  the lack of inversion symmetry and the strong spin-orbit coupling which are present in the material. Skyrmions also appeared in the relativistic context in the Skyrme work \cite{Sky}, in the search for a unified theory of mesons and baryons. In the present work, however, we focus on vortices in the planar relativistic Maxwell-Higgs model, that is, we revisit the Nielsen and Olesen work \cite{NO}, but we enlarge the system to accommodate extra degrees of freedom. We do this in the next Sec. \ref{sec:model}, where we study two distinct models. In Sec. \ref{sec:end} we end the study discussing other possibilities related to the results of this work.

\section{Models}
\label{sec:model}

\subsection{First model, with $U(1) \times Z_2$ symmetry}

We start with the following Lagrange density, with dimensionless quantities and metric tensor $\eta_{\mu\nu}$ such that ${\rm diag}(\eta_{\mu\nu})=(1,-1,-1)$, with $\mu,\nu=0,1,2$,
\be\label{lvortex}
{\cal L}=-\frac14 f(\chi)F_{\mu\nu}F^{\mu\nu}\!\!+\!|D_\mu\varphi|^2\!+\!\frac12\partial_\mu \chi \partial^\mu\chi-V(|\varphi|,\chi).
\ee
In this model, $\chi$ is a neutral real scalar field, $\varphi$ is a charged complex scalar field, $F_{\mu\nu}=\partial_\mu A_\nu-\partial_\nu A_\mu$, $D_\mu=\partial_\mu+iA_\mu$, and $A_\mu$ is a vector field. $V(|\varphi|, \chi)$ is the potential, which is supposed to account for interactions between the neutral and charged fields. Also,
$f(\chi)$ is a nonnegative real function of the neutral field, and for some constant values of $\chi$, the model leads us back to the standard Maxwell-Higgs system, with $V(|\varphi|)$ being the Higgs-type potential. The symmetry in this case is $U(1)\times Z_2$, accounting for the local $U(1)$ symmetry and the global $Z_2$ symmetry which is governed by the real scalar field $\chi$. To search for vortices we consider static fields, take $A_0=0$ and
\be\label{ansatz1}
\chi=\chi(r),\;\;\; \varphi=g(r)e^{in\theta},\;\;\; \vec{A}=-\frac{\hat\theta}{r} (a(r)-n),
\ee
where $n$ is a nonvanishing integer that represents the vorticity and the functions must obey the boundary conditions $\chi(0)=\chi_0$, $\chi(\infty)=\chi_\infty$, $g(0)=0$, $g(\infty)=1$, $a(0)=n$ and $a(\infty)=0$. In this case the magnetic field is given by $B=-a^\prime/r$, and the flux is quantized, $\Phi=n\,\Phi_0$, with $\Phi_0=2\pi$ being the basic flux. Note that we are taking unity electric charge and we will also use $n=1$, for simplicity.

The equations of motion are written as
\bes\label{secansatz}
\begin{align}
\frac{1}{r} \left(r \chi^\prime\right)^\prime &= f_\chi\frac{{a^\prime}^2}{2r^2} + V_\chi,\\ 
\frac{1}{r} \left(r g^\prime\right)^\prime &= \frac{a^2g}{r^2} + \frac12 V_g,\\
r\left(f\frac{a^\prime}{r}\right)^\prime &= 2ag^2,
\end{align}
\ees
where $f_\chi=df/d\chi$, $V_\chi = \partial V/\partial\chi$, and $V_{|\vphi|}=\partial V/\partial|\vphi|$. The presence of nonlinearity in the potential adds further nonlinearities in the above equations of motion, which is mandatory to obtain stable localized structures. The static configurations governed by the above equations of motion engender energy density
\be\label{rhoans}
\rho= f(\chi) \frac{{a^\prime}^2}{2r^2} +  {g^\prime}^2 + \frac{a^2g^2}{r^2} + \frac12{\chi^\prime}^2 + V(g,\chi).
\ee
The above field configurations are invariant under rotations in the plane, and one may follow the lines of Ref.~\cite{N3,V1} to find a first order framework that minimizes the energy of our system. It arises with the potential
\be\label{potgen}
V(|\vphi|,\chi) =  \frac{1}{2}\frac{\left(1-|\vphi|^2\right)^2}{f(\chi)} + \frac12\frac{W^2_\chi}{r^2},
\ee
where $W=W(\chi)$ is an auxiliary function that controls the neutral field. The presence of the radial coordinate in the last term of the potential was first studied in \cite{PRL}, and is important to support stable planar kinklike solution to be constructed from the neutral field $\chi$. With the above potential, the energy that comes from Eq.~\eqref{rhoans} is minimized to
$E=2\pi+ 2\pi\left|W(\chi(\infty))-W(\chi(0))\right|$ if the following first order equations are satisfied
\be\label{fochi}
\chi^\prime = \pm\frac{W_\chi}{r},
\ee
and
\be\label{fov}
g^\prime = \pm\frac{ag}{r},\;\;\;\;\;\;\;\;
-\frac{a^\prime}{r} =\pm \frac{\left(1-g^2\right)}{f(\chi)}.
\ee
The positive and negative signs in Eq.~\eqref{fochi} are related by the change $r\to1/r$, and the ones in Eqs.~\eqref{fov} by $a\to-a$, so we only consider the positive signs. Since the first order equations provide minimum energy solutions, they are stable against small fluctuations of the field configurations, because the fluctuations cannot decrease the energy of the minimum energy solutions. 

The importance of the first order equations \eqref{fochi} and \eqref{fov} goes beyond their simplicity; since they ensure minimum energy to the corresponding solutions, they are then linearly stable, discarding the necessity to conduct hard linear stability analysis for the allowed vortex configurations. Another interesting result is that the energy which is explicitly written just above Eq. \eqref{fochi} does not depend on $f(\chi)$, so we can change the allowed forms of $f(\chi)$ without modifying the energetic behavior of the solutions. And yet, another property of the vortex configurations that solve the first order equations concerns their topological behavior. To see how this works, let us introduce the topological current
\be
j_T^\mu=\frac12\epsilon^{\mu\sigma\delta} F_{\delta\sigma},
\ee
where $\epsilon^{\mu\sigma\delta}$ represents the Levi-Civita symbol. It is conserved, such that $\partial_\mu j_T^{\mu}=0$: the topological charge can be written in the form $Q_T=2\pi (a(0)-a(\infty))$, using \eqref{ansatz1}; thus, the boundary conditions on $a(r)$ lead us with
$ Q_T=2\pi$. We then see that the function $f(\chi)$ does not modify neither the total energy nor the topological charge of the vortex solutions that obey the first order equations \eqref{fochi} and \eqref{fov}. Another property of interest is that the magnetic field is such that its flux gives $\Phi=2\pi$, and so coincides with the topological charge. In the present context one notices that although the function $f(\chi)$ controls the magnetic field, it does not interfere in its flux. Thus, in summary the function $f(\chi)$ can be used to distribute the magnetic field and the enrgy density in the plane without altering neither its flux nor the total energy and stability of the vortex solution. These are important properties of the above model and its
companion first order equations.
		\begin{figure}[t!]
		\centering
		\includegraphics[width=4.2cm]{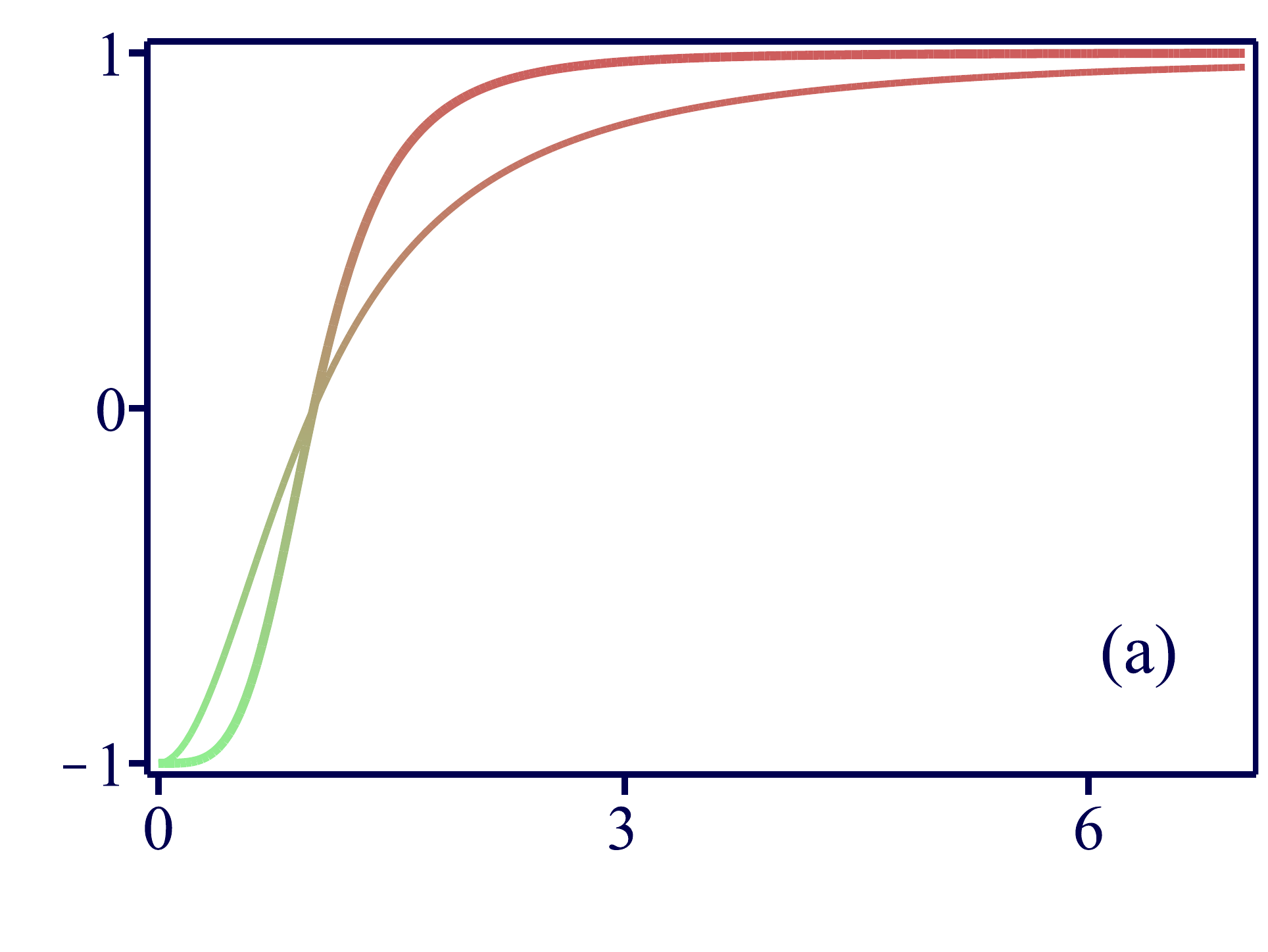}
		\includegraphics[width=4.2cm]{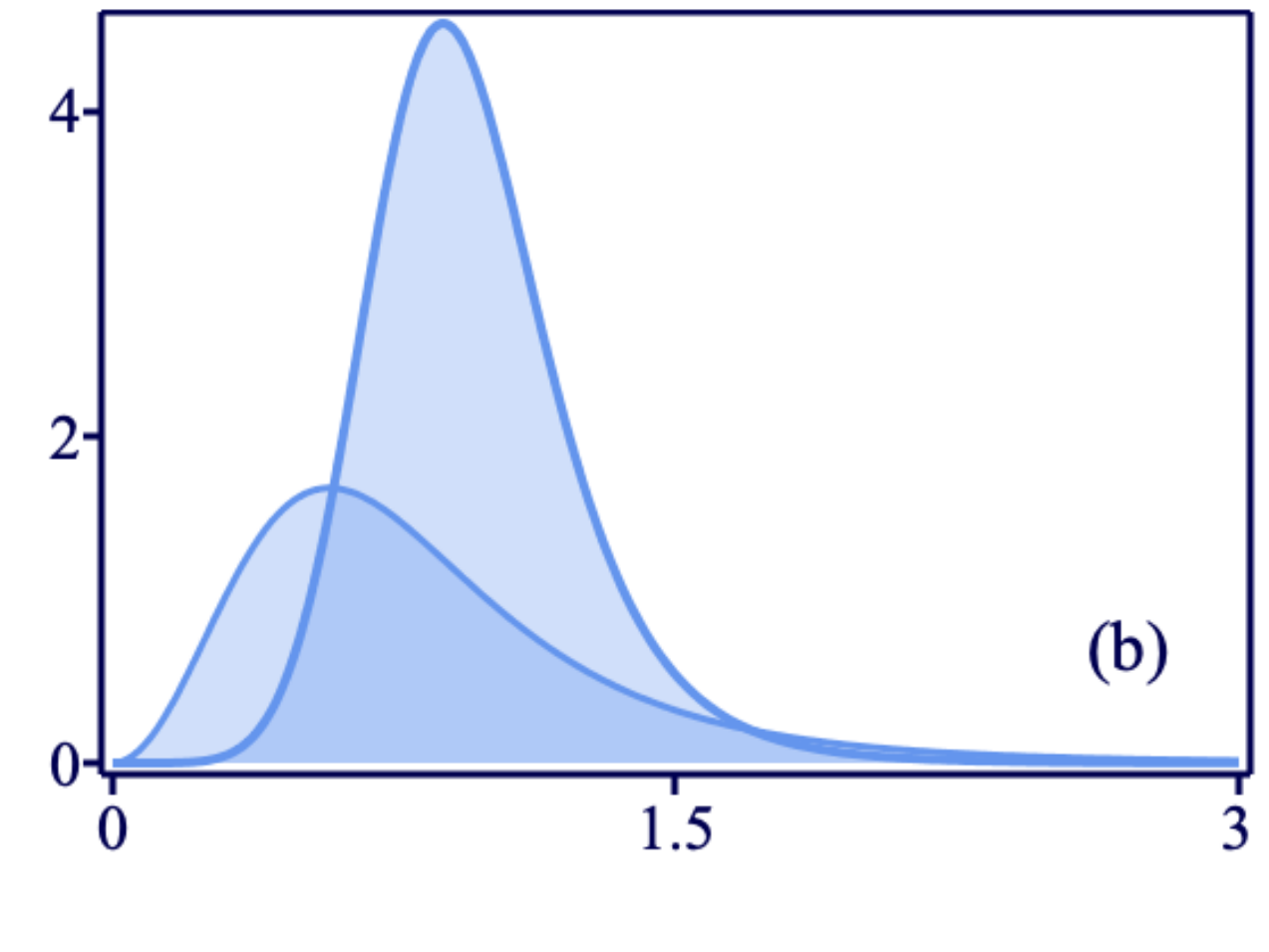}
		\caption{The first model. The kinklike solution $\chi(r)$ (a) and its energy density $\rho_\chi(r)$ (b) are depicted for the cubic case, with $r_0=1$, and $\alpha=1$ and $2$. The thickness of the lines increases with $\alpha$.}
		\label{figv1}
		\end{figure}

To find solutions, however, one must suggest the form of $W(\chi)$ to determine the kinklike solution in Eq.~\eqref{fochi} to be used to model the function $f(\chi)$ in order to feed the first order equations \eqref{fov} to describe the vortex. Before going into this, we notice that solutions of the above first order equations allow us to write the energy density \eqref{rhoans} as two separated contributions, in the form
$\rho=\rho_{vor}+ \rho_\chi$, where
\be\label{rhos}
\rho_{vor} = f(\chi) \frac{{a^\prime}^2}{r^2} + 2{g^\prime}^2,\;\;\;\;\;\;\;\; \rho_\chi = {\chi^\prime}^2.
\ee
To emphasize the role played by the several fields, we remark that the $\chi$ field which is guided by the $Z_2$ symmetry has to be capable of generating a localized structure to entrap the vortex described by the $\varphi$ and $A_\mu$ fields that are guided by the local $U(1)$ symmetry. In this sense, the potential to describe the $\chi$ field has to have at least two minima, say $\pm\bar{\chi}$, for which $V(|\varphi|,\pm \bar{\chi})=V(|\varphi|)$, leading us back to the Maxwell-Higgs potential. If we further require that in the limit $\chi\to\pm\bar{\chi}$, the model becomes the Maxwell-Higgs model that supports the standard Nielsen-Olesen vortex configuration, this imposes that the function $f(\pm\bar\chi)$ becomes the same positive constant, and this further constrains the function $f(\chi)$.

Let us now concentrate on the construction of explicit models. We consider two distinct possibilities, with the $\chi$ field engendering distinct nonlinearities of current interest, to see how the nonlinearity associated with the $Z_2$ symmetry may modify the shape of the vortex described by the local $U(1)$ symmetry.

\subsubsection{Cubic nonlinearity}

Let us investigate the case in which the $\chi$ field engenders cubic nonlinearity in the equation of motion. For the model under investigation, this is implemented with the choice $W(\chi) = \alpha \chi-\alpha\chi^3/3$, such that
\be\label{chi4}
W_\chi=\alpha(1-\chi^2),
\ee
which vanishes at the values $\pm1$, determining the minima and the values $\chi_0$ and $\chi_\infty$ to be used to define the solution
$\chi(r)$. The kinklike solution that appears from Eq.~\eqref{fochi} with the positive sign is
\be\label{solvchi}
\chi(r) = \frac{r^{2\alpha}-r_0^{2\alpha}}{r^{2\alpha}+r_0^{2\alpha}},
\ee
where $r_0$ is a parameter associated to the size of the kink and $\alpha$ controls the slope of the solution at $r=r_0$. In Fig.~\ref{figv1}, we depict this solution and its respective energy density $\rho_\chi(r)$ for $r_0=1$ and $\alpha=1$ and $2$.

		\begin{figure}[t!]
		\centering
		\includegraphics[width=4.2cm]{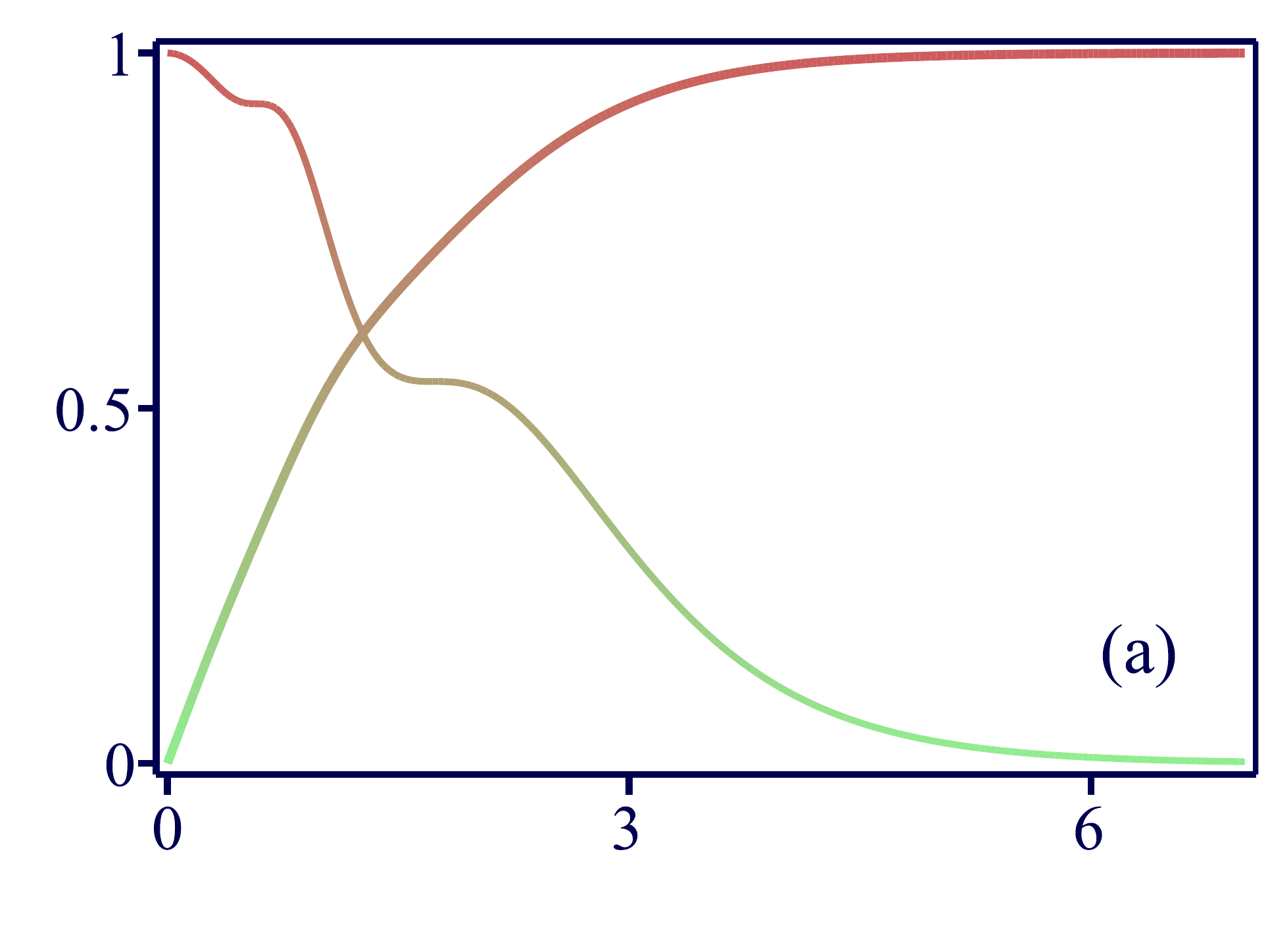}
		\includegraphics[width=4.2cm]{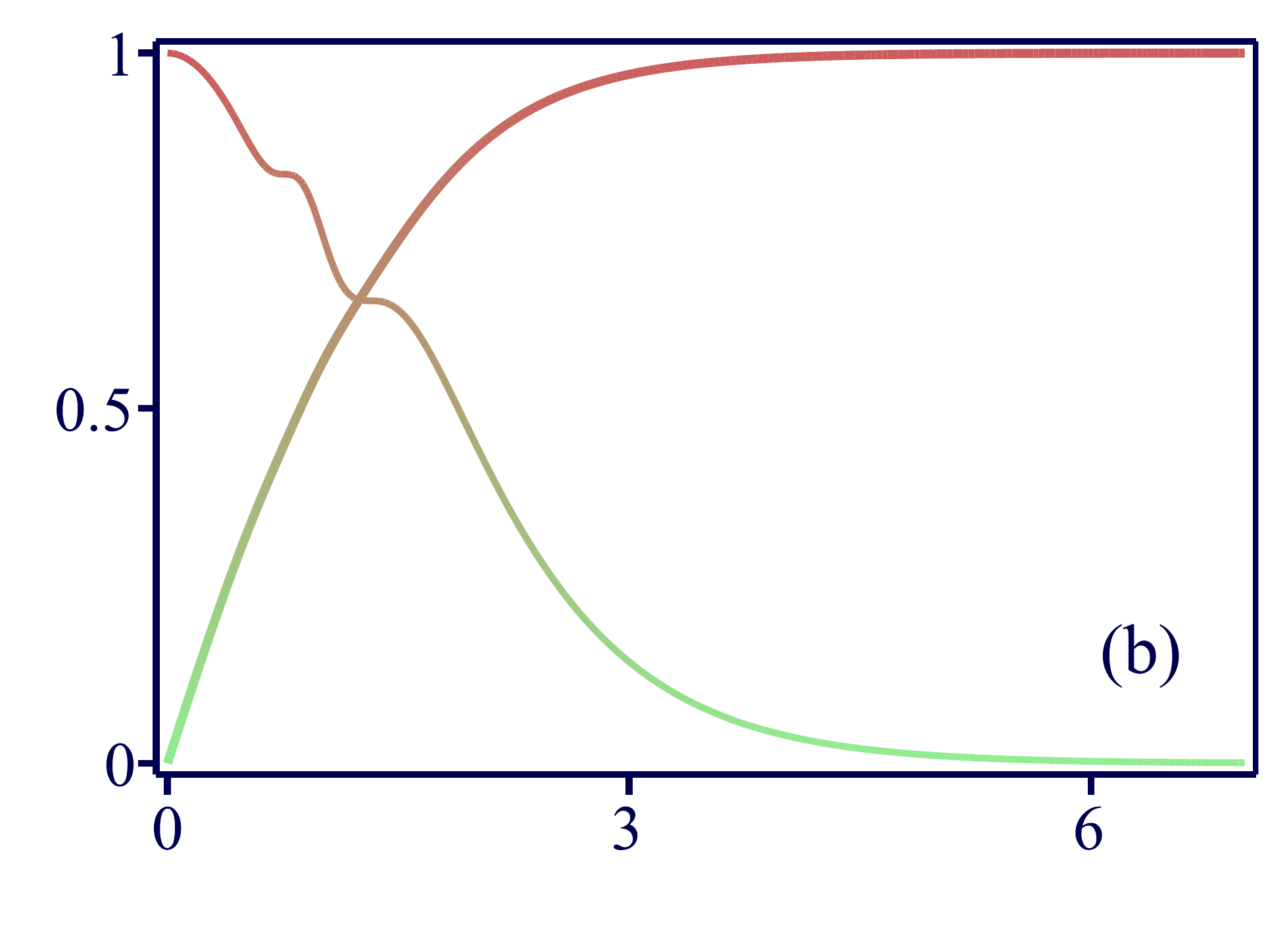}
		\includegraphics[width=4.2cm]{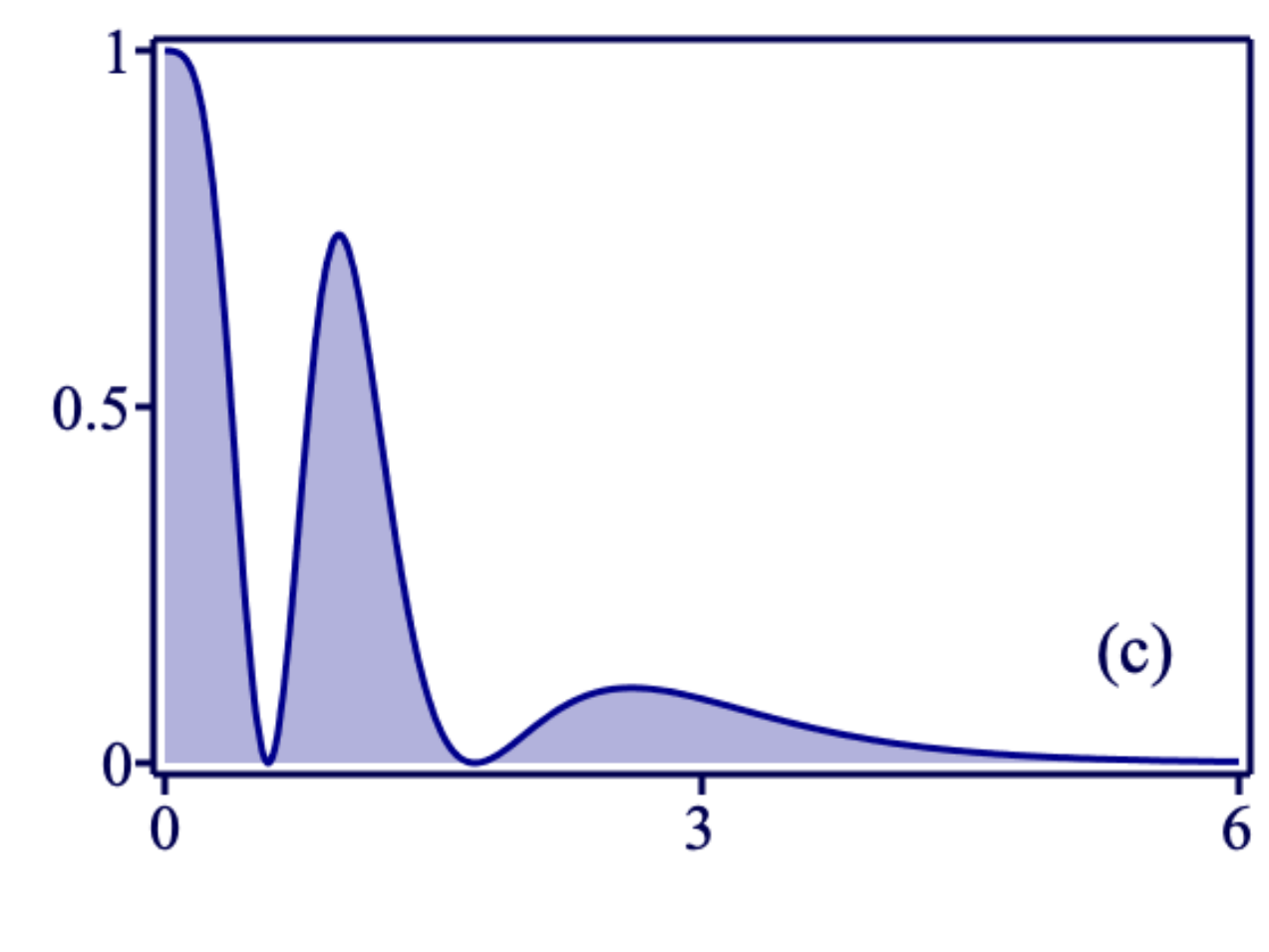}
		\includegraphics[width=4.2cm]{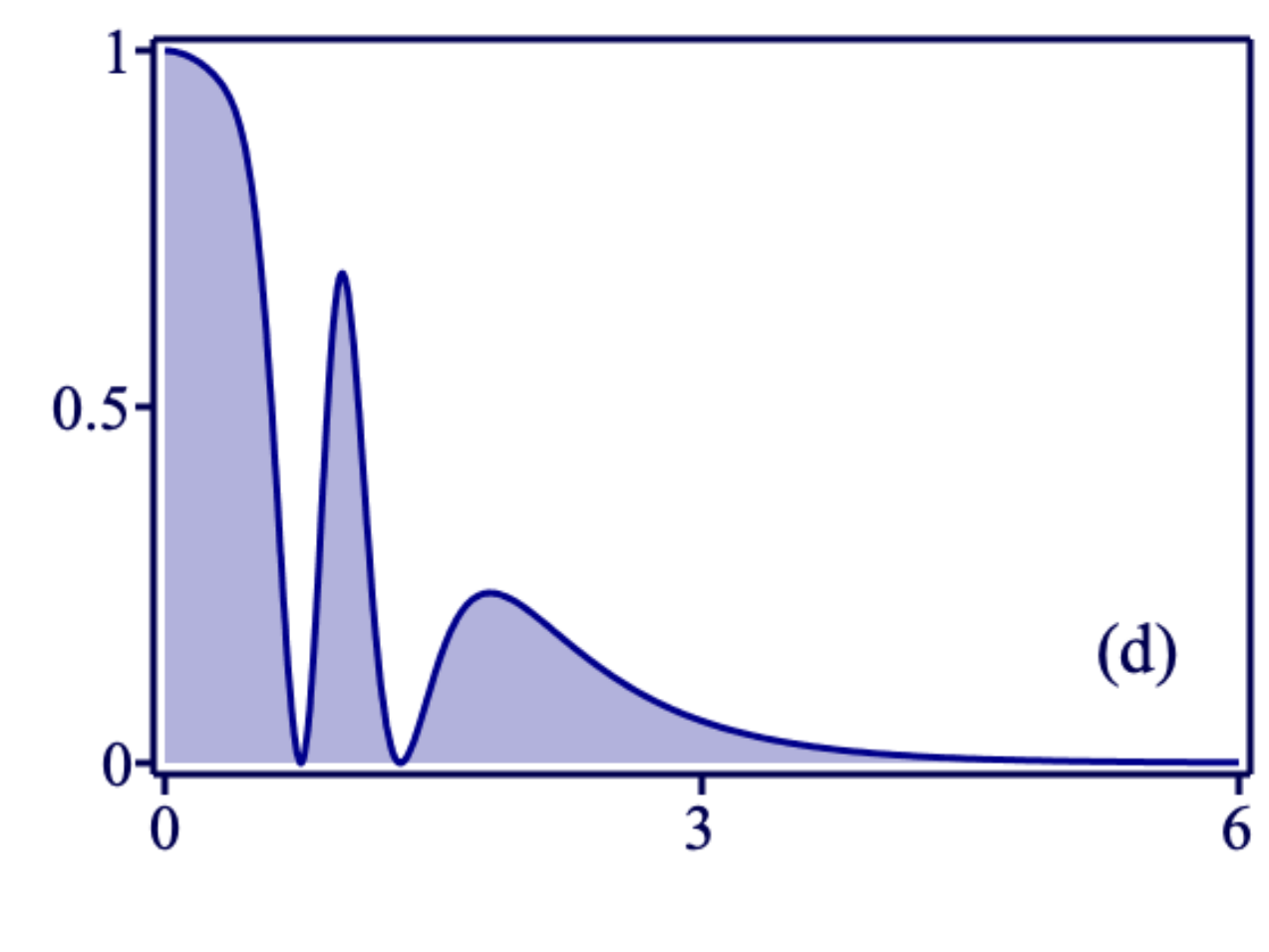}
		\includegraphics[width=4.2cm]{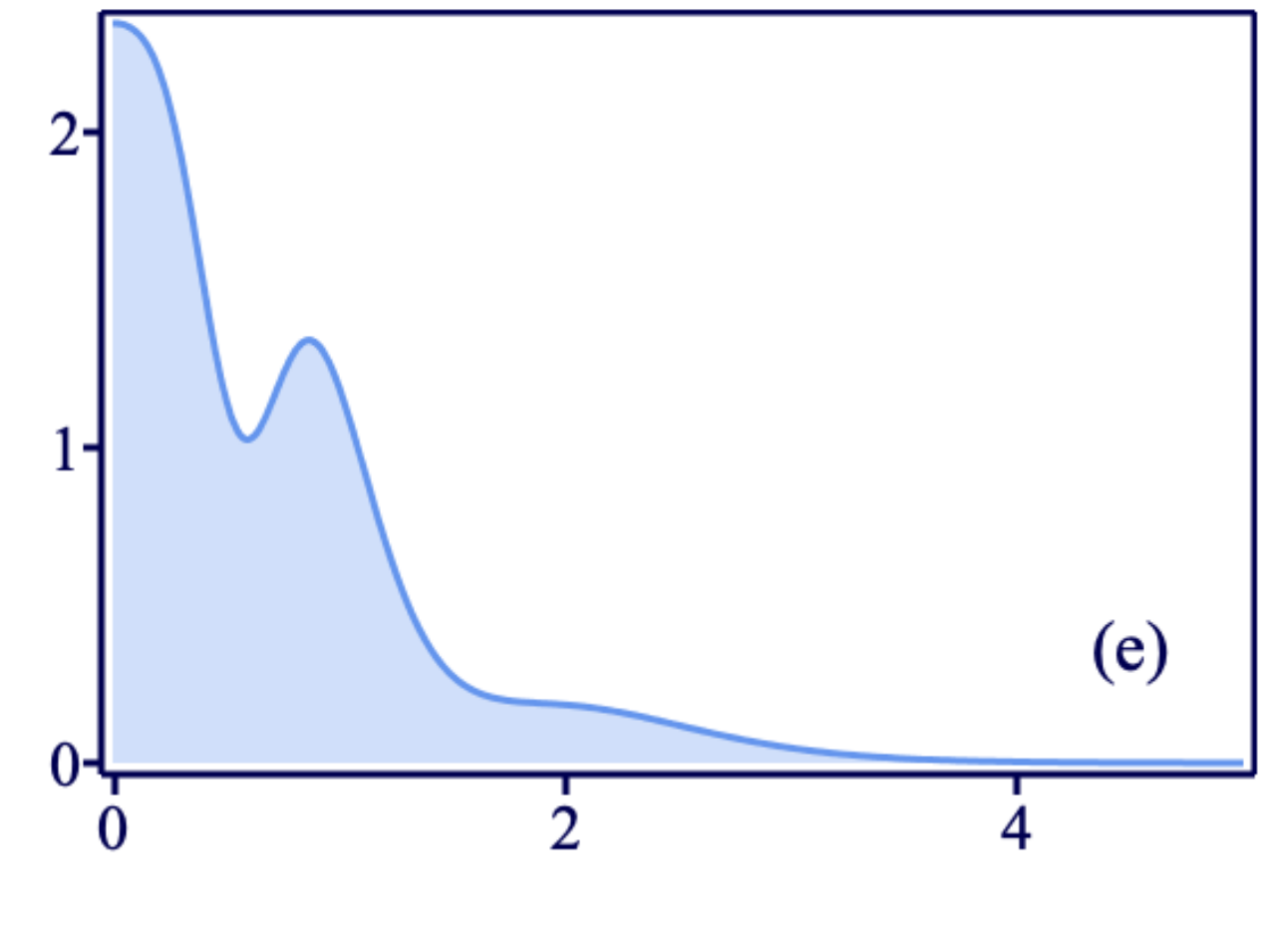}
		\includegraphics[width=4.2cm]{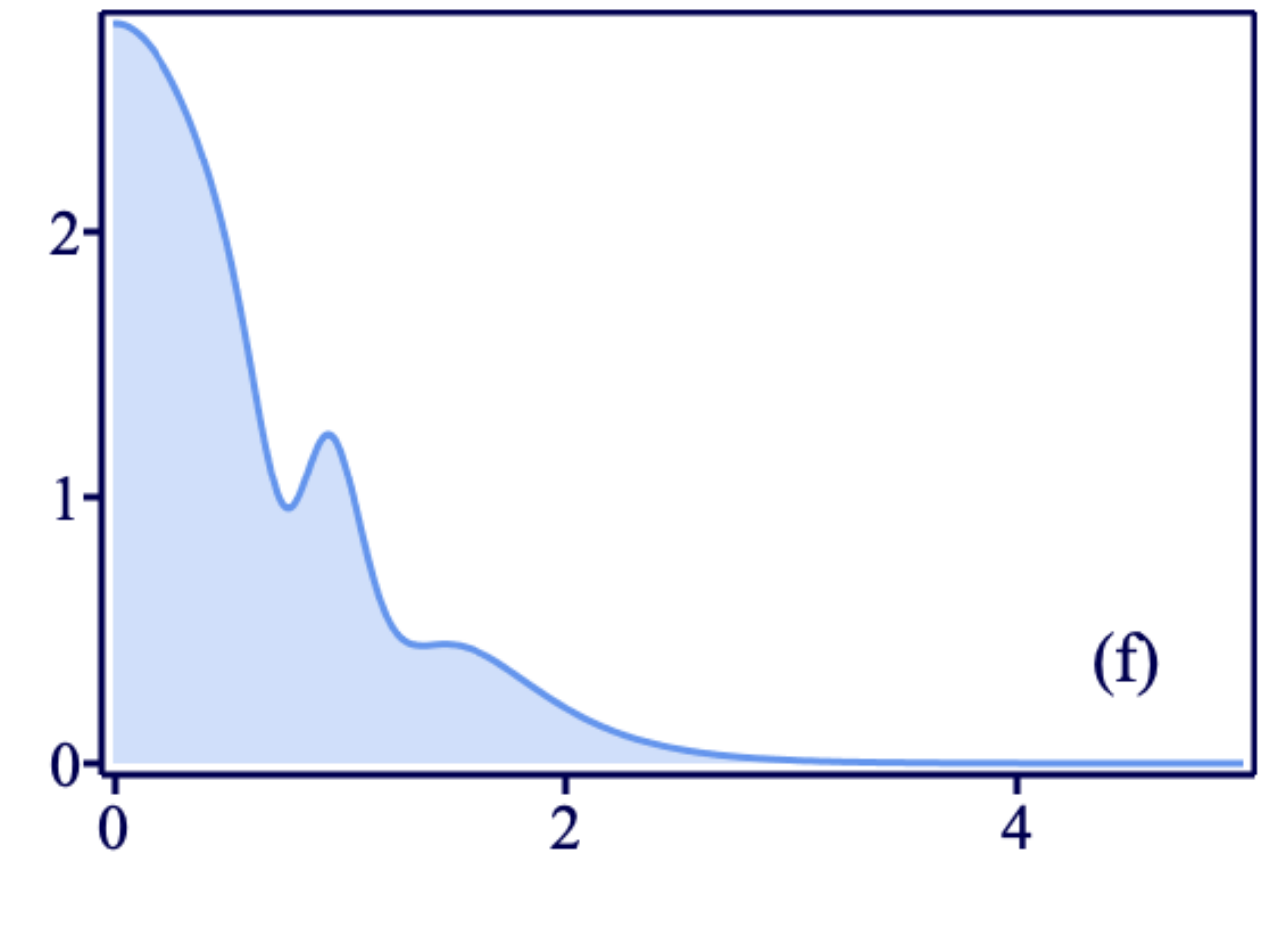}
		\caption{The first model. The vortex solutions $a(r)$ (descending line) and $g(r)$ (ascending line) in the panels (a) and (b), the magnetic field $B(r)=-a^\prime/r$ ((c) and (d)) and the energy density $\rho_{vor}(r)$ ((e) and (f)). They are depicted for the cubic case, with $n=1$, $r_0=1$, $m=1$ and $\alpha=1$ (left) and $2$ (right), respectively.}
		\label{figv2}
		\end{figure}

\begin{figure}[t!]
		\centering
		\includegraphics[width=4.2cm]{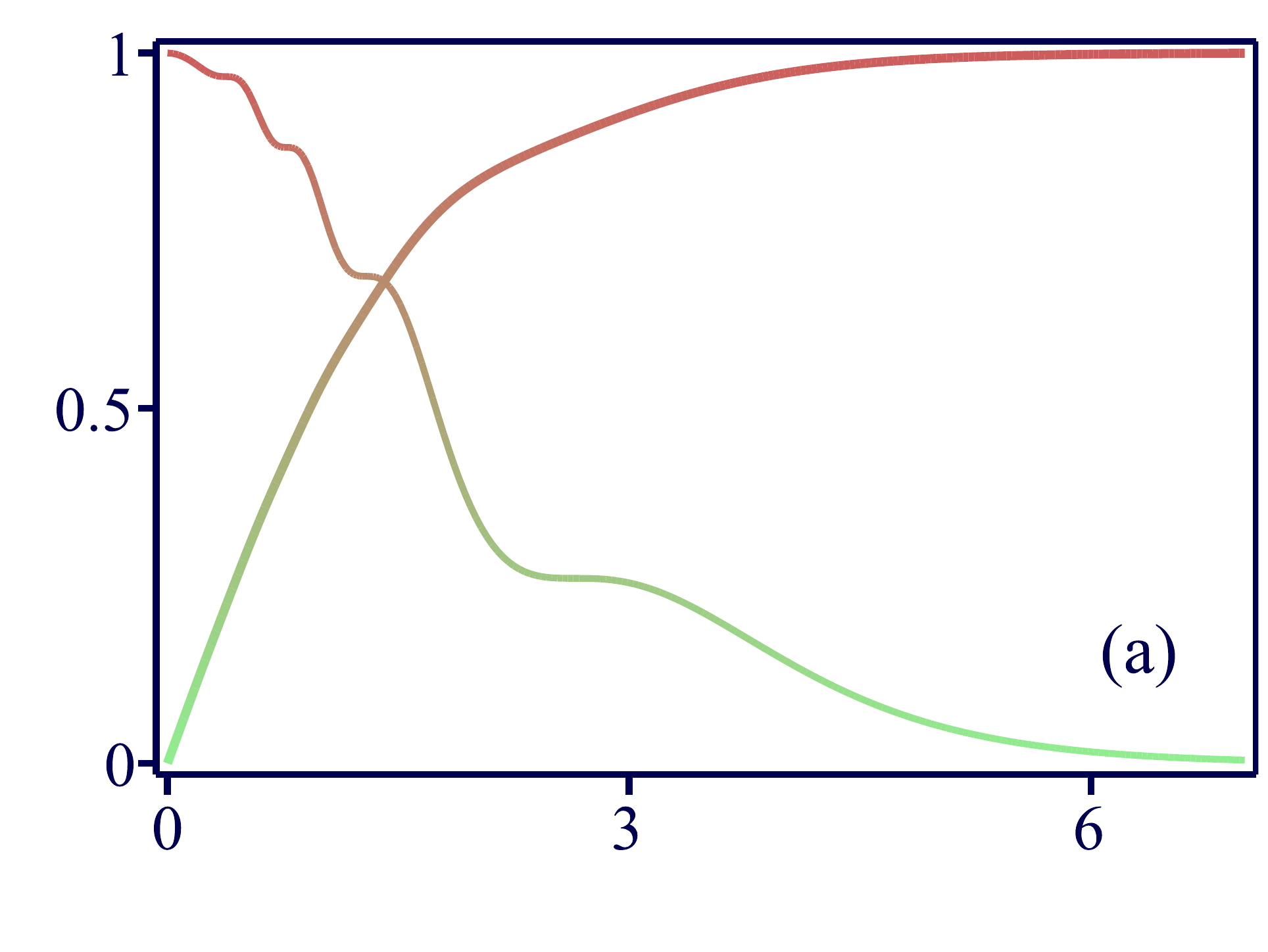}
		\includegraphics[width=4.2cm]{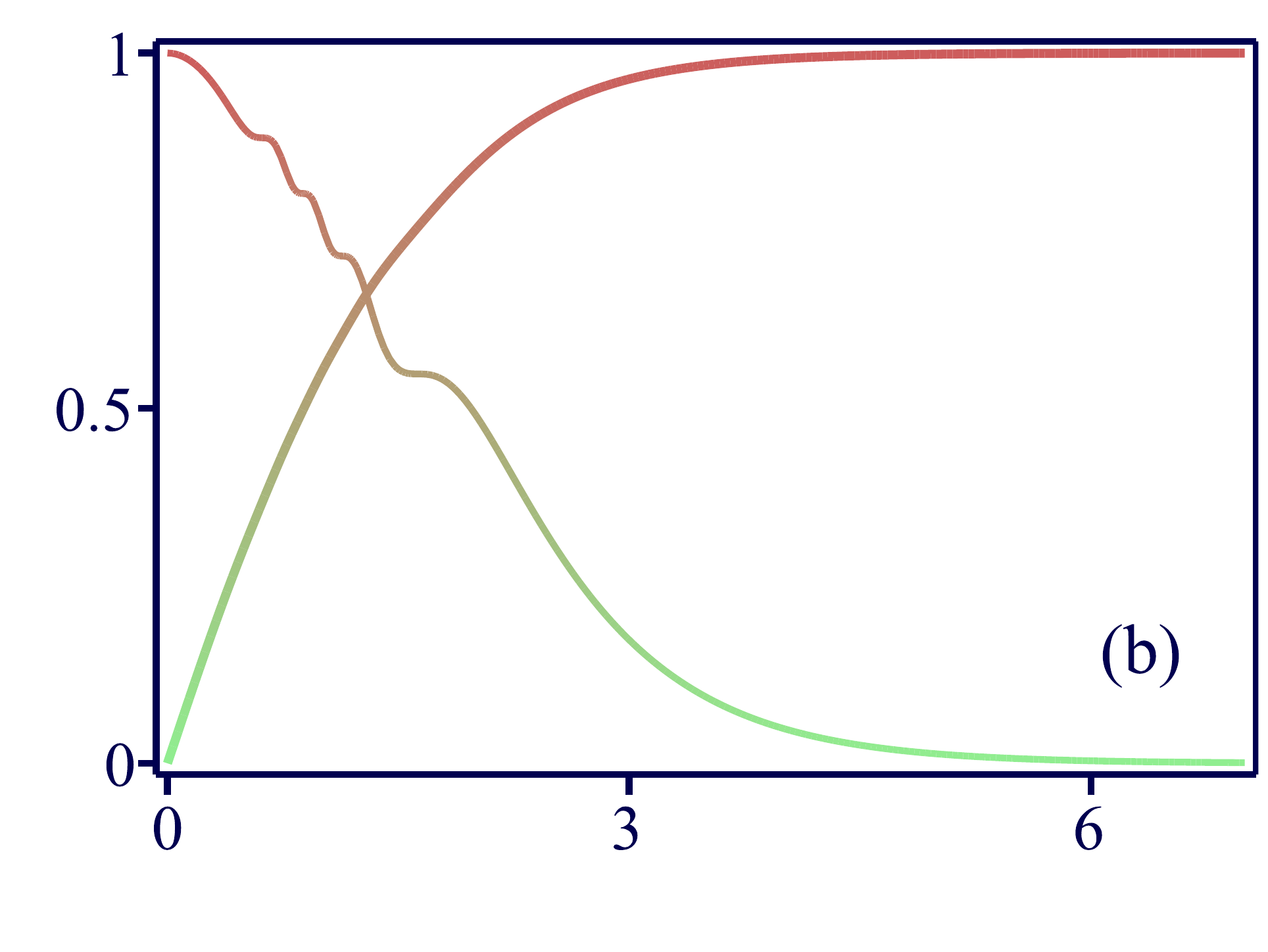}
		\includegraphics[width=4.2cm]{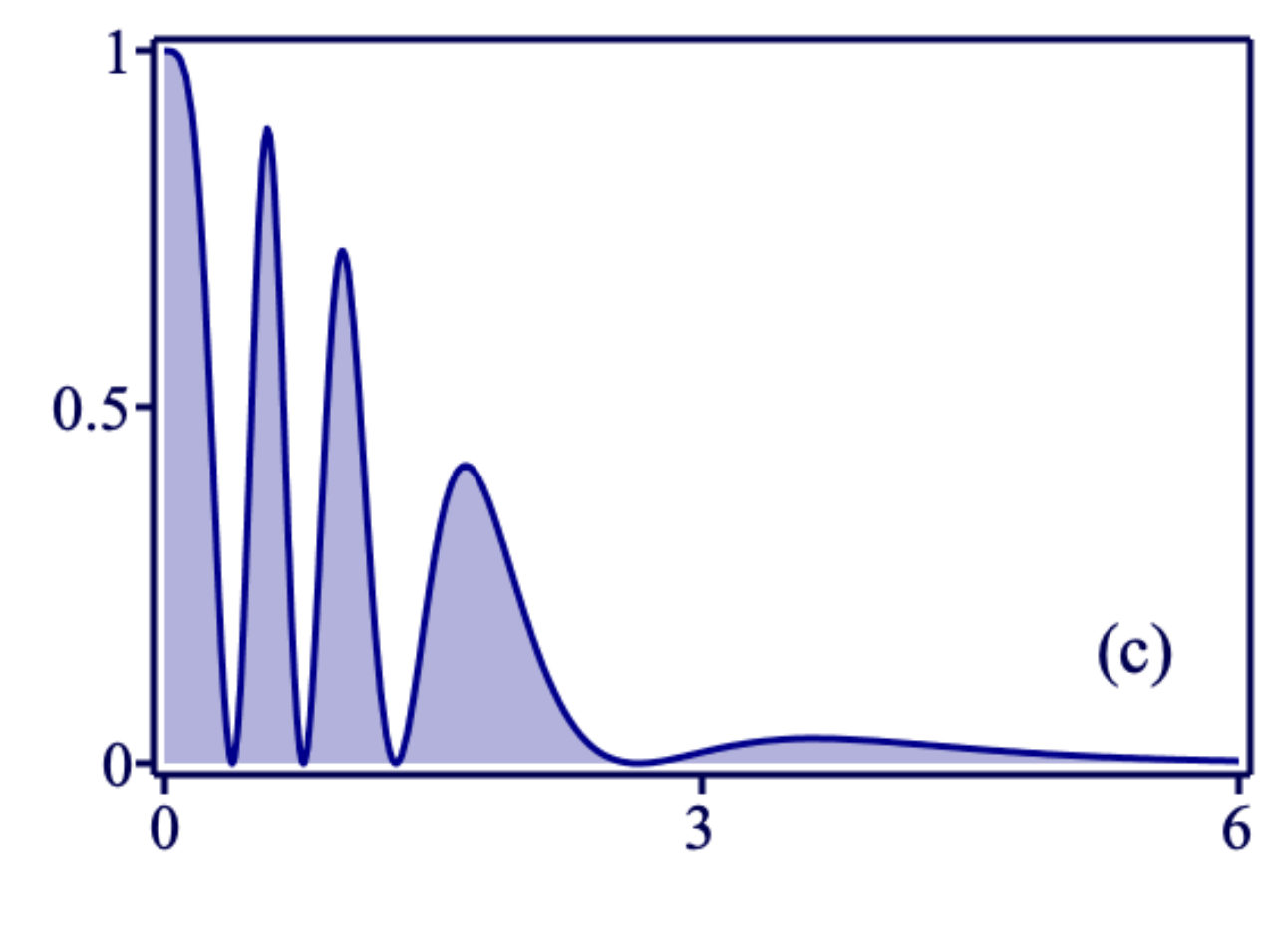}
		\includegraphics[width=4.2cm]{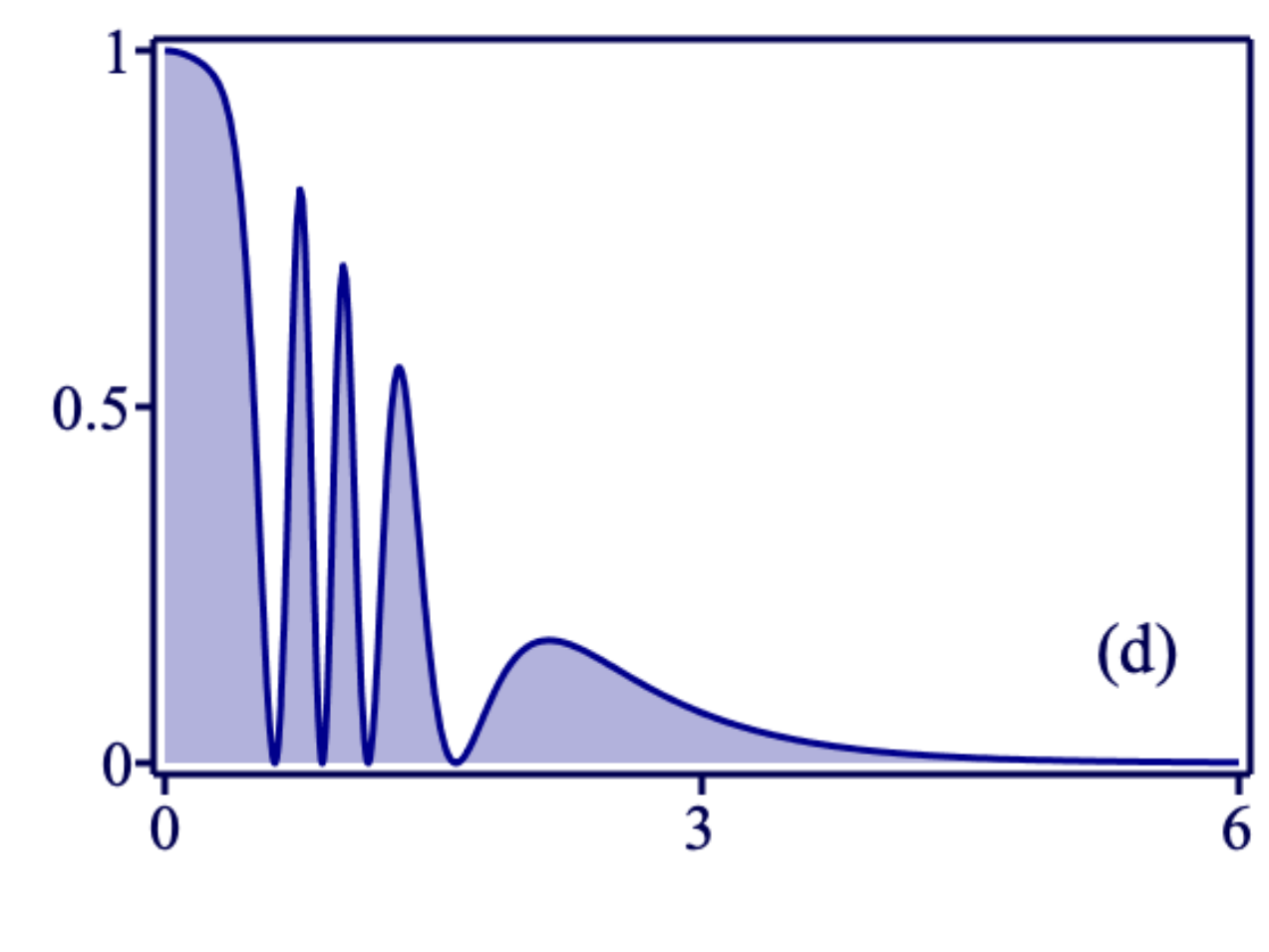}
		\includegraphics[width=4.2cm]{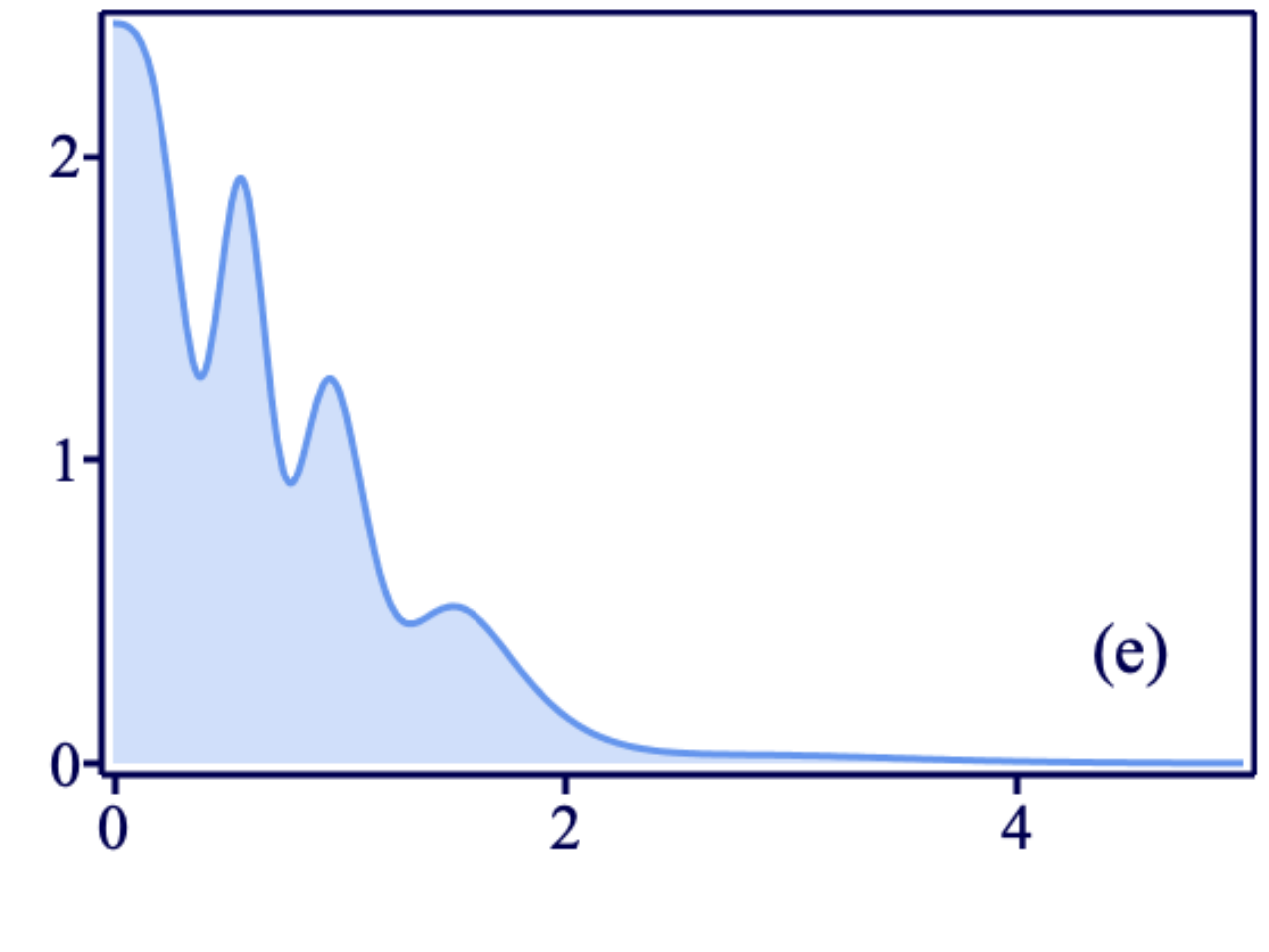}
		\includegraphics[width=4.2cm]{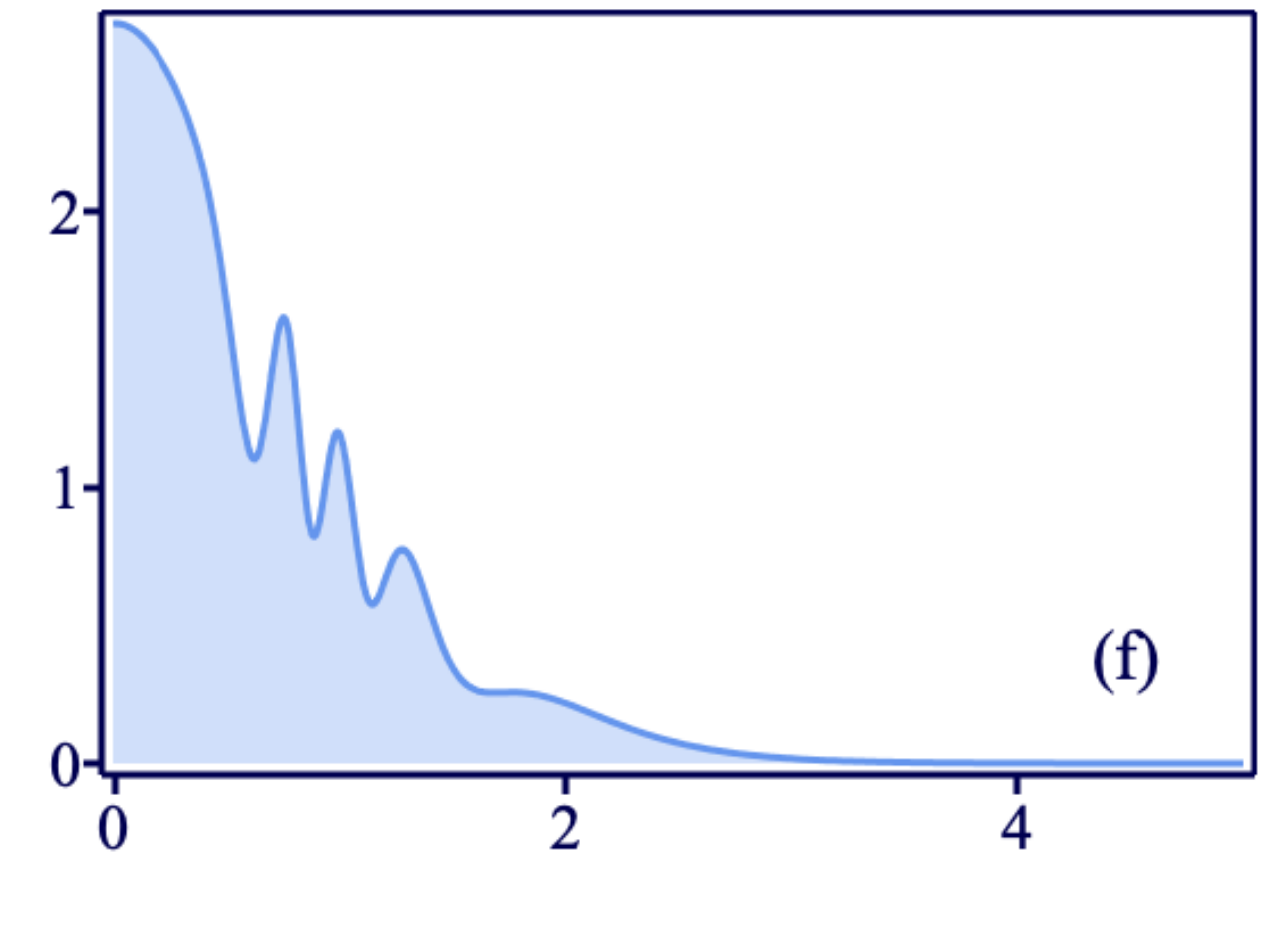}
		\caption{The first model. The vortex solutions $a(r)$ (descending line) and $g(r)$ (ascending line) in the panels (a) and (b), the magnetic field $B(r)=-a^\prime/r$ ((c) and (d)) and the energy density $\rho_{vor}(r)$ ((e) and (f)). They are depicted for the cubic case, with $n=1$, $r_0=1$, $\lambda=0$, $m=2$ and $\alpha=1$ (left) and $2$ (right), respectively.}
		\label{figv3}
		\end{figure}

\begin{figure}[t!]
		\centering
		\includegraphics[width=4cm]{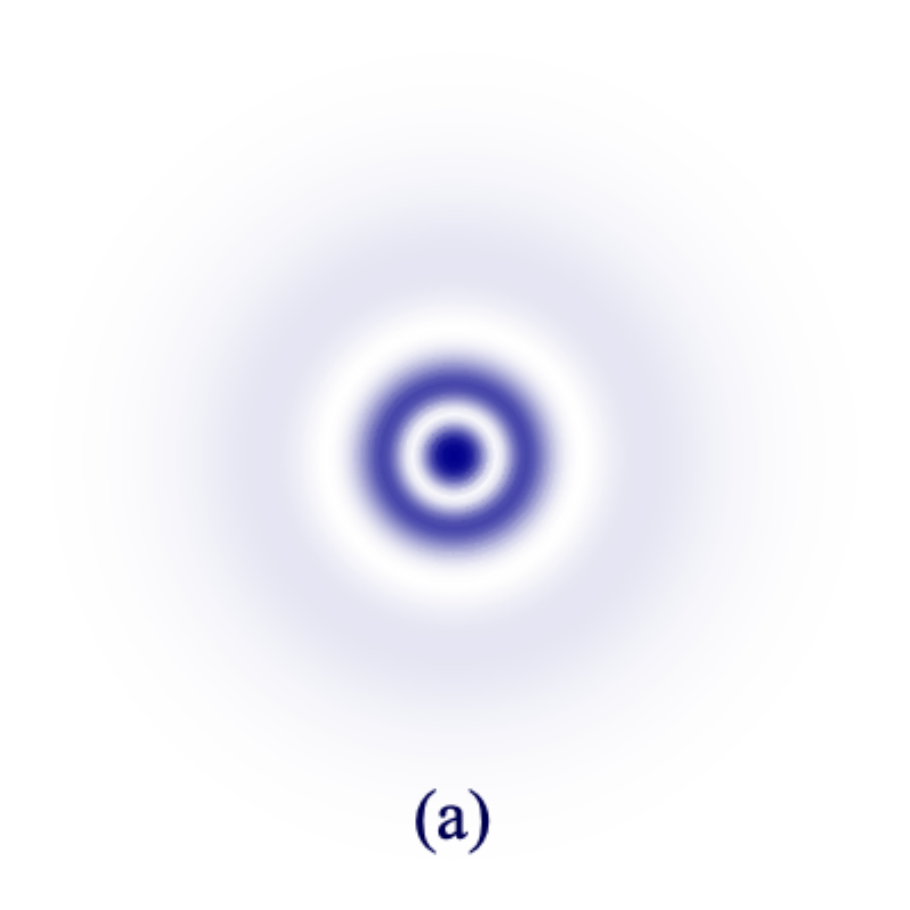}
		\includegraphics[width=4cm]{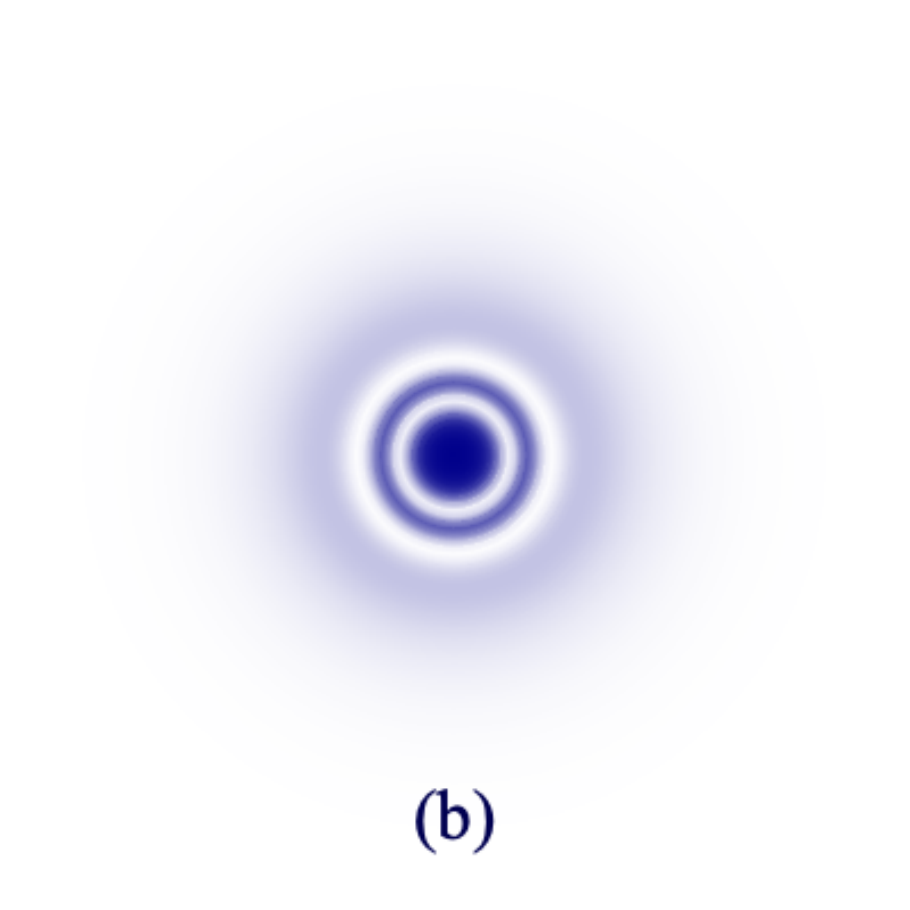}
		\includegraphics[width=4.0cm]{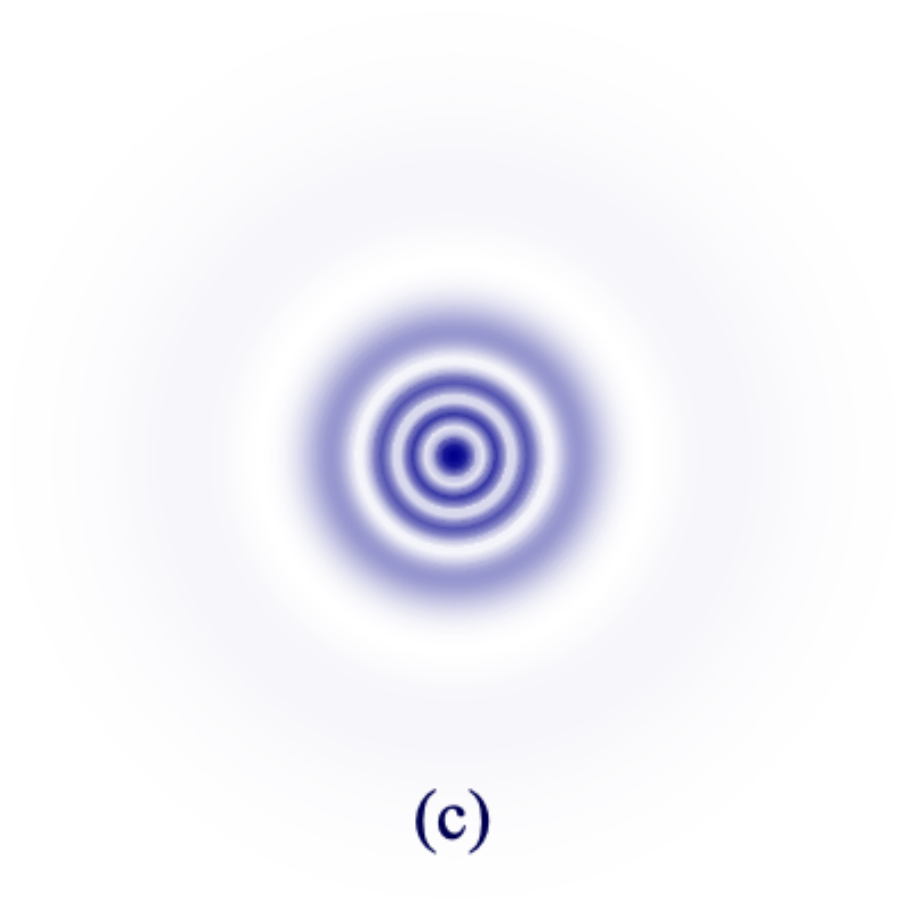}
		\includegraphics[width=4.0cm]{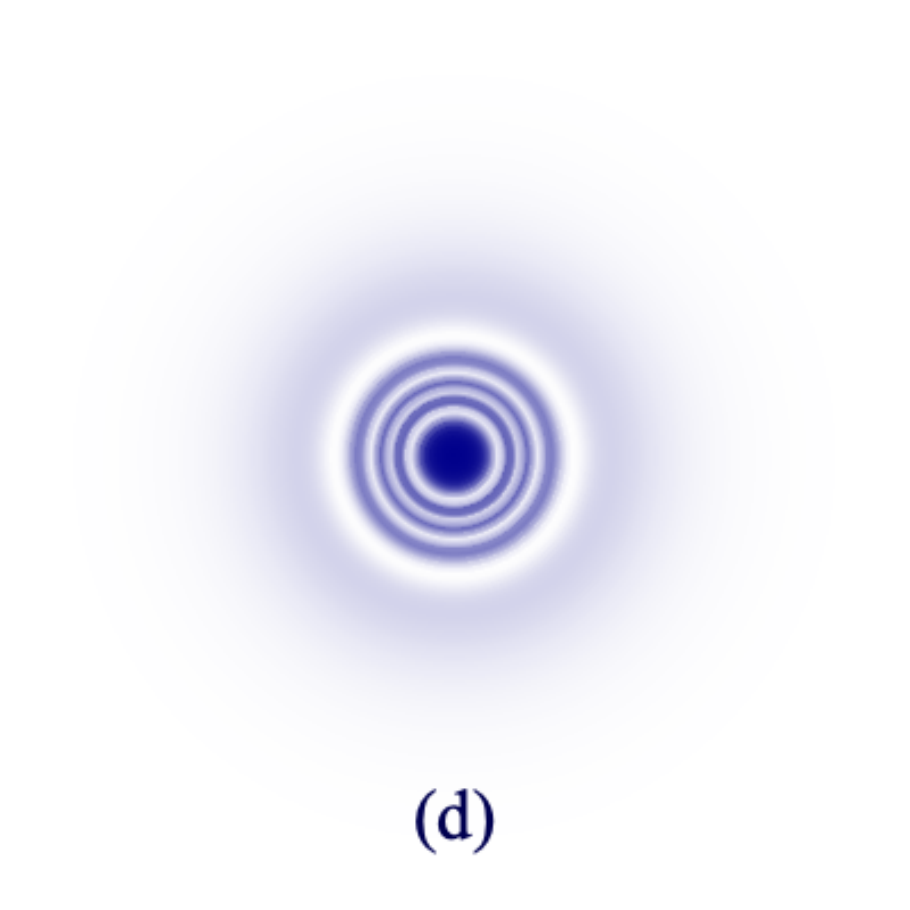}
		\caption{The first model. The magnetic field associated to the vortex is depicted in the plane for the cubic case, with $n=1$, $r_0=1$, $\lambda=0$, $\alpha=1$ and $m=1$ (a), $\alpha=2$ and $m=1$ (b), $\alpha=1$ and $m=2$ (c), and $\alpha=2$ and $m=2$ (d).}
		\label{figv4}
		\end{figure}

We use the above solution as a source for the function that controls the magnetic permeability, which we take as $f(\chi)=(1+\lambda^2)/(\lambda^2+\cos^2(m\pi\chi))$, $m\in\mathbb{N}$ and $\lambda \in \mathbb{R}$. This is of practical interest because for large values of $\lambda$, the function $f(\chi)$ tends to unity, the $\chi$ field decouples, and $g(r)$ and $a(r)$ become the Nielsen-Olesen vortex configurations. For $\lambda$ small, the model supports novel configurations, which we investigate below. We first consider the case
$\lambda=0$. Here, the first-order equations \eqref{fov} with the upper sign become
\be\label{fovmult}
g^\prime =\frac{ag}{r},\;\;\;\;\;
-\frac{a^\prime}{r} = \cos^2\!\left(\!m\pi\,\frac{r^{2\alpha}-r_0^{2\alpha}}{r^{2\alpha}+r_0^{2\alpha}}\right)\!\left(1-g^2\right).
\ee
We solve these equations numerically, and in Fig.~\ref{figv2} we depict the vortex solutions, magnetic field $B=-a^\prime/r$ and energy density $\rho_{vor}(r)$ in Eq.~\eqref{rhos} for $n=1$ and for $r_0=1$, $m=1$ and $\alpha=1$ and $2$, from which we see how the parameter $\alpha$ modifies the vortex. In order to verify the role of the parameter $m$, we also depict the aforementioned quantities for $n=1$, and for $r_0=1$, $m=2$, and $\alpha=1$ and $2$ in Fig.~\ref{figv3}. We further illustrate the model depicting in Fig.~\ref{figv4} the magnetic fields in the plane for $n=1$, and for $r_0=1$ and some values of $m$ and $\alpha$. We see that the magnetic field of the vortex engender substructures associated to these parameters: a single central disk and $2m$ external rings, with the radius of the central disk increasing with increasing
$\alpha$. The energy density also shows similar internal structure, but in a more subtle manner.


\begin{figure}[t!]
		\centering
		\includegraphics[width=4.2cm]{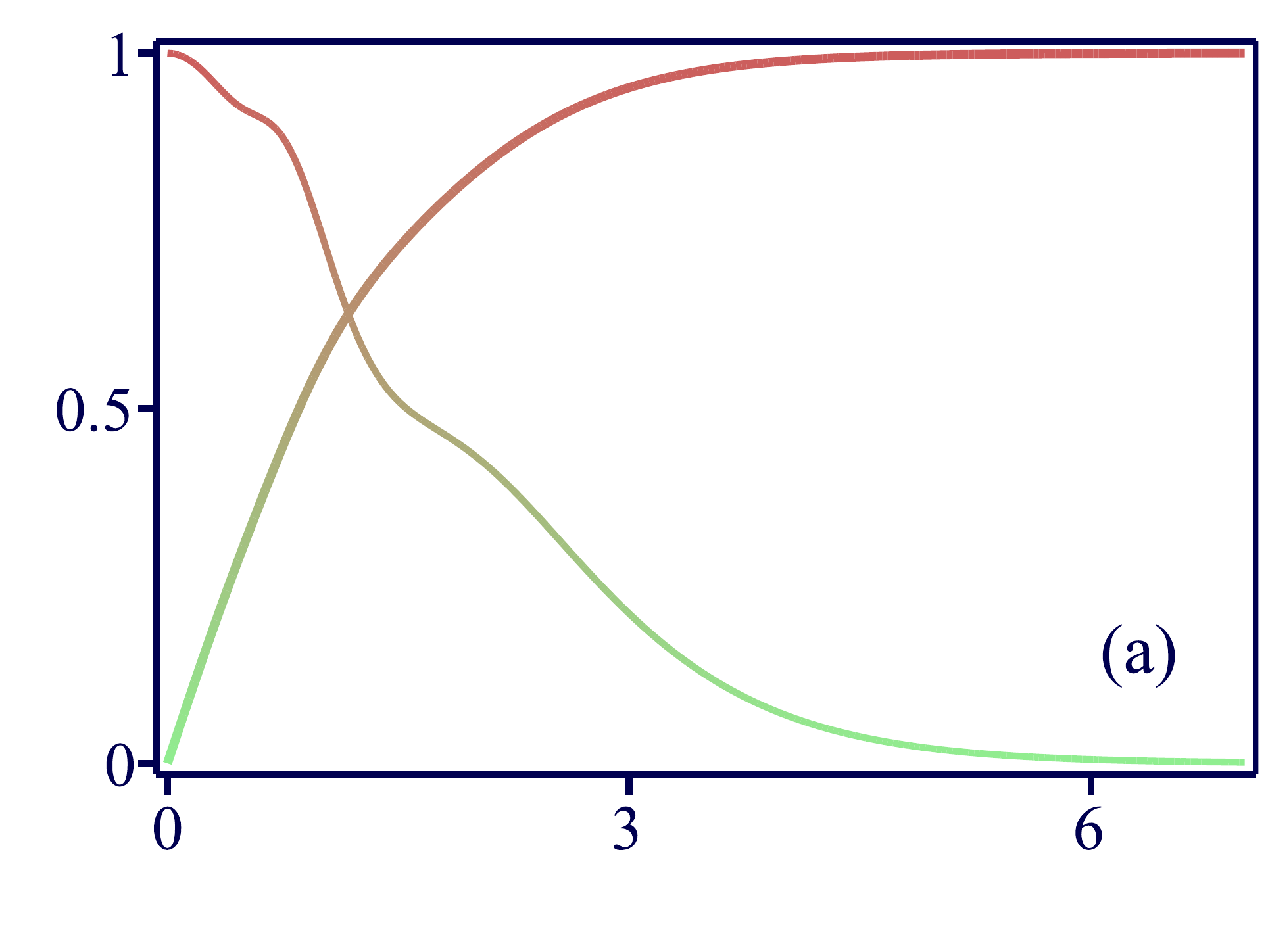}
		\includegraphics[width=4.2cm]{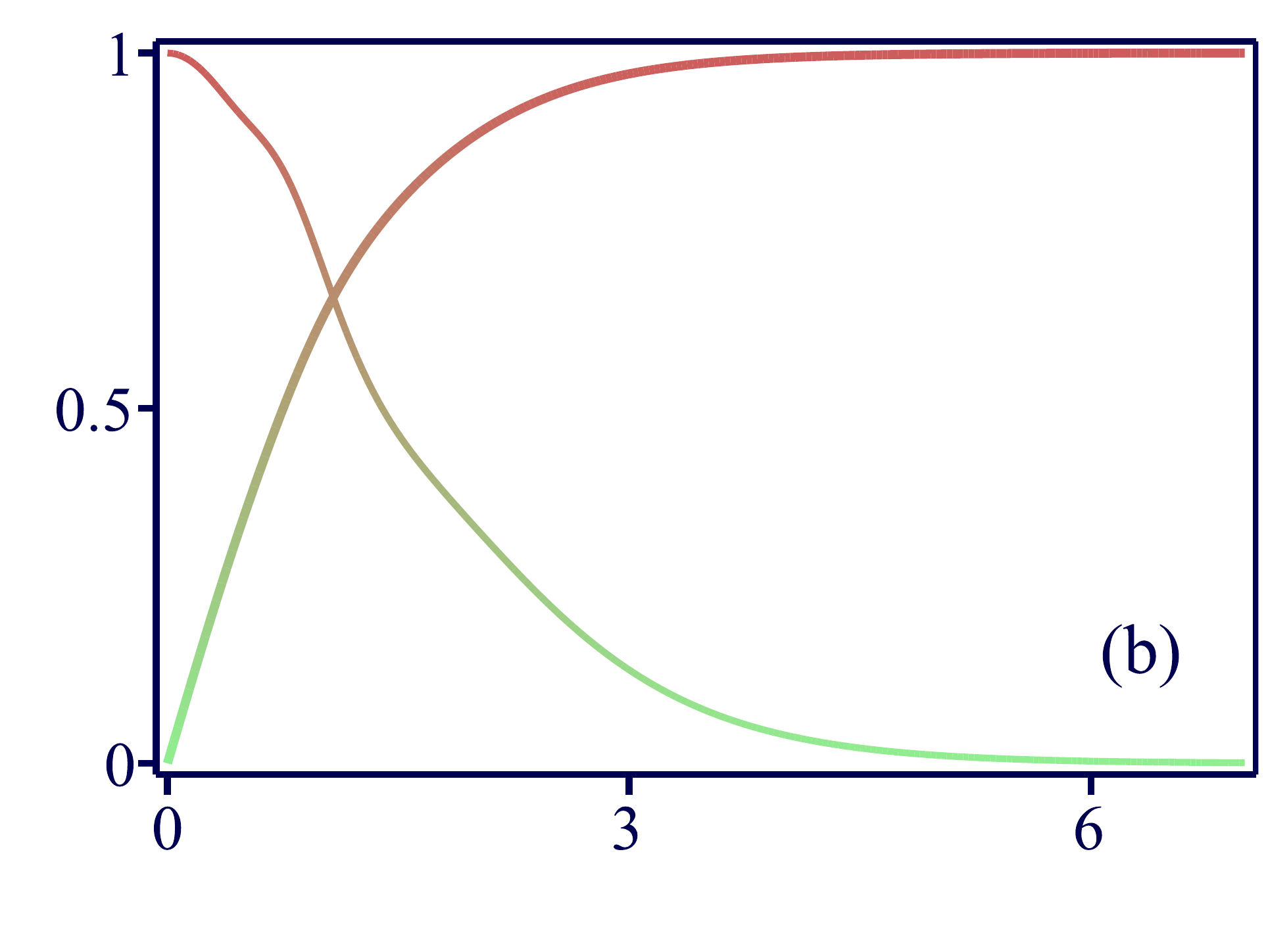}
		\includegraphics[width=4.2cm]{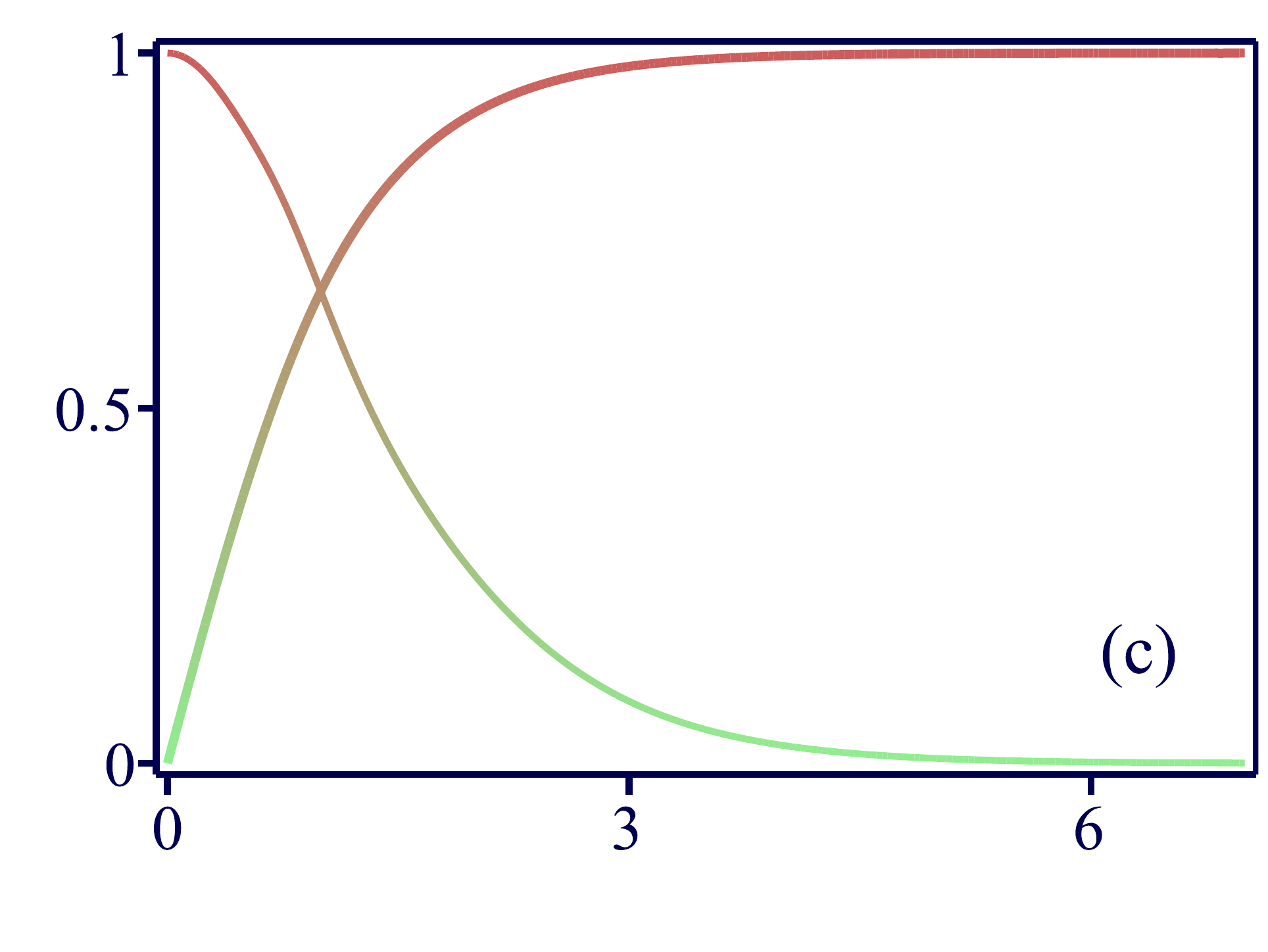}
		\includegraphics[width=4.2cm]{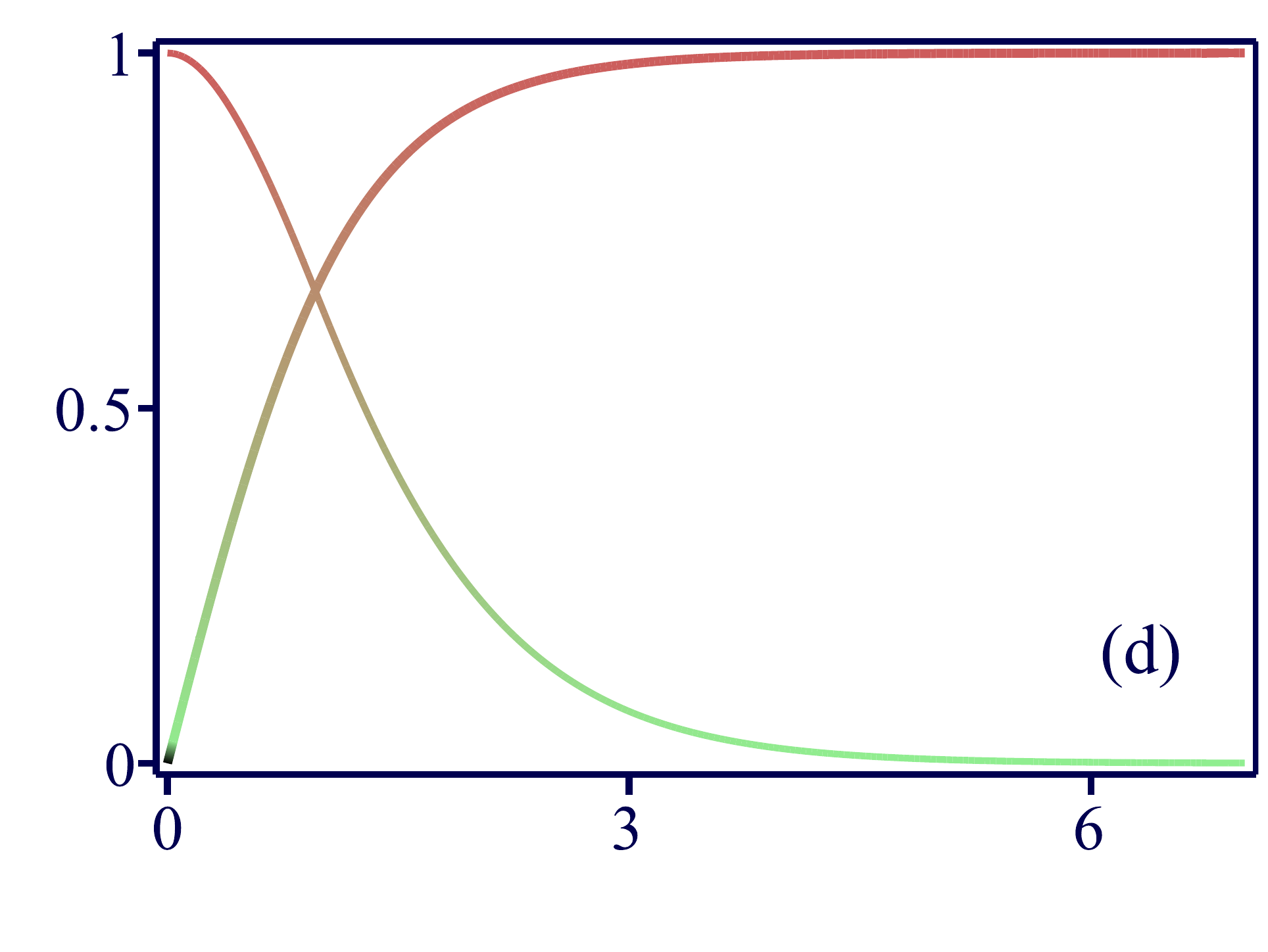}
		\includegraphics[width=4.2cm]{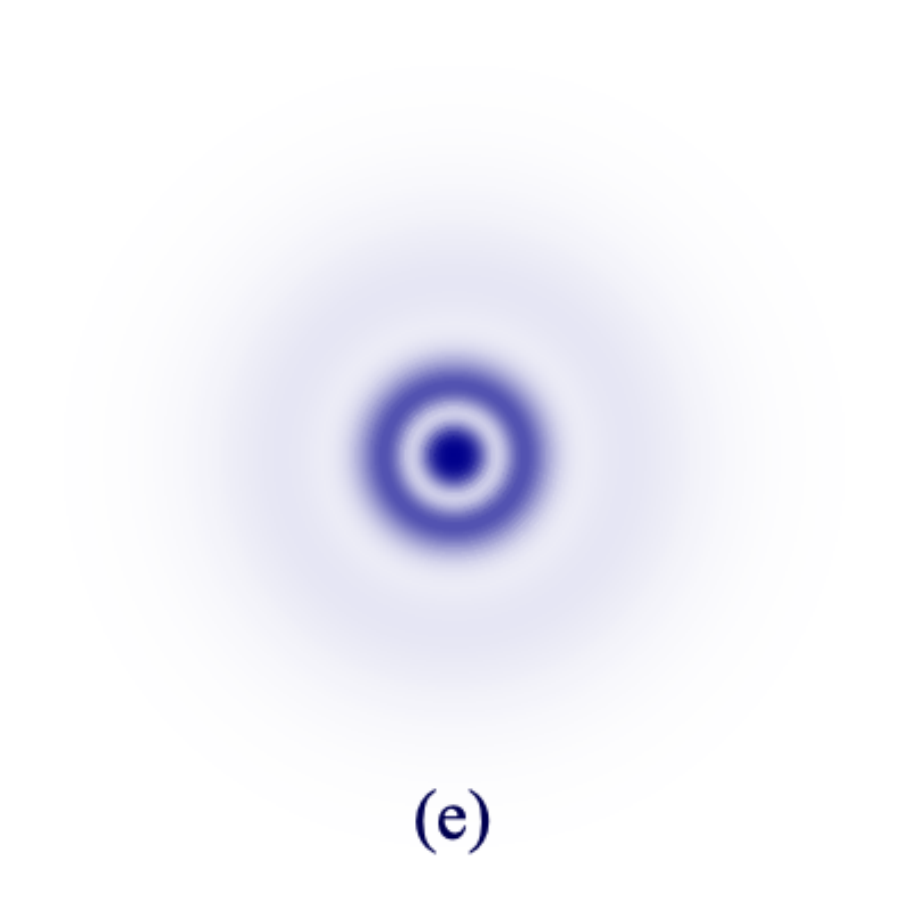}
		\includegraphics[width=4.2cm]{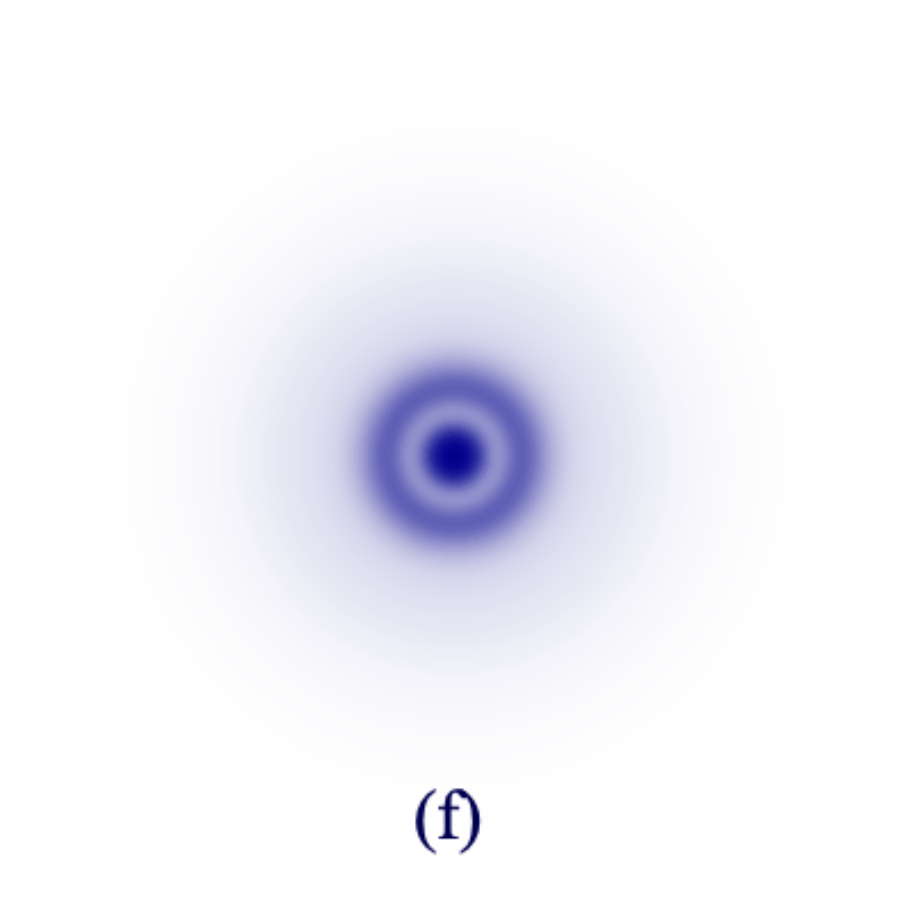}
		\includegraphics[width=4.2cm]{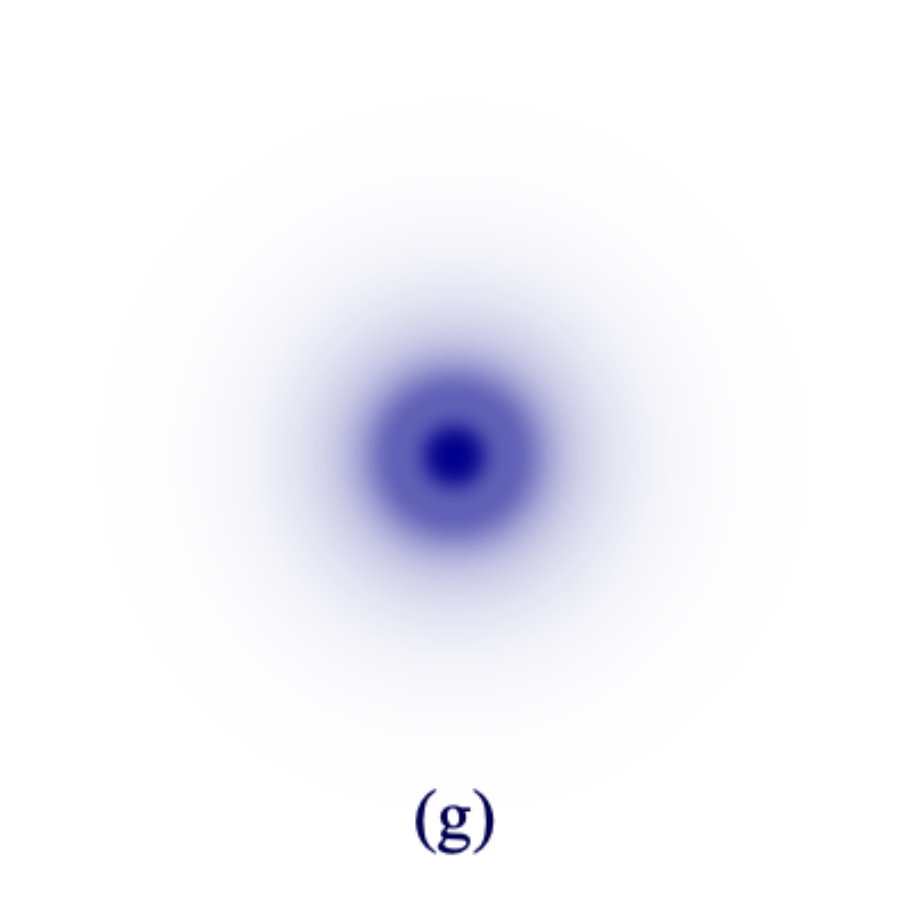}
		\includegraphics[width=4.2cm]{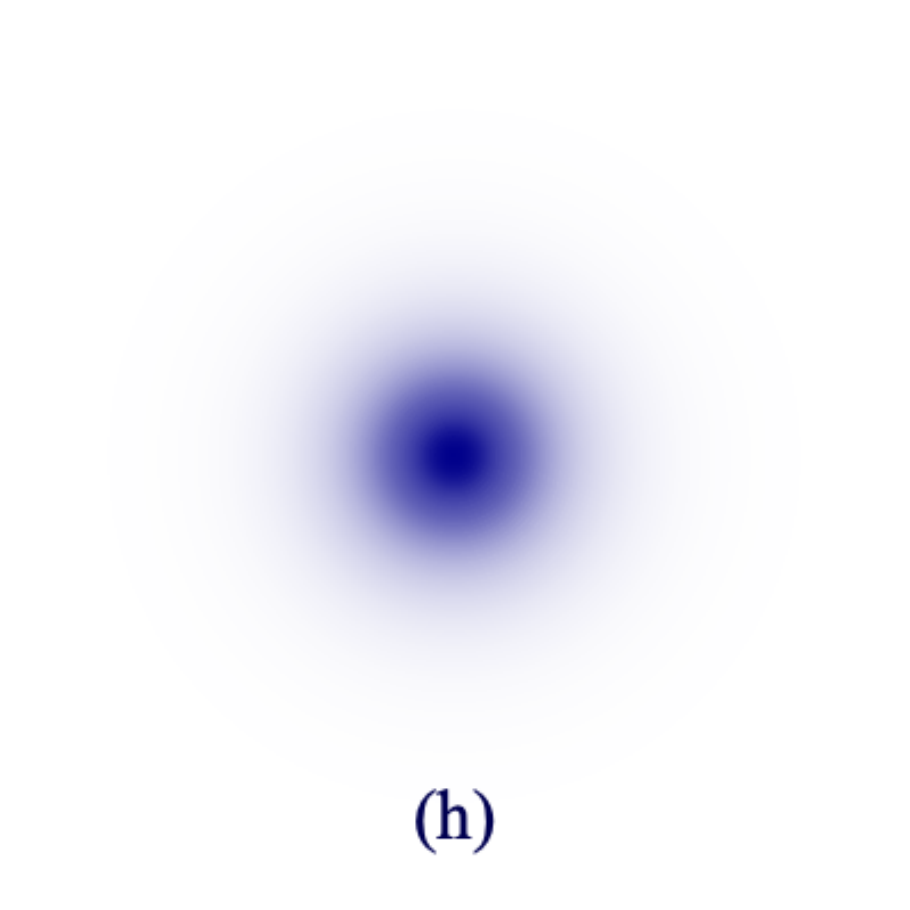}
		\caption{The first model. The vortex solutions $a(r)$ (descending line) and $g(r)$ (ascending line) (four top panels), and the magnetic field associated to the vortex, which is depicted in the plane (four bottom panels). We take $n=1$, $r_0=1$, $m=1$, $\alpha=1$, and $\lambda=0.5$ ((a) and (e)), $1$ ((b) and (f)), $2$ ((c) and (g)), and $4$ ((d) and (h)).}
		\label{figv5}
		\end{figure}

We now consider the case of a nonvanishing $\lambda$. The investigation is similar to the previous one, so in Fig. \ref{figv5} we only depict $g(r)$, $a(r)$ and the magnetic field in the plane for $n=1$ and $r_0=1$, $m=1,$ $\alpha=1$, and $\lambda=0.5, 1,2$ and $4$. We remind that the case of $\lambda=0$ is in the top left of Fig. \ref{figv4}, and the case $\lambda=4$ in the bottom right of
Fig. \ref{figv5} is essentially the magnetic field of the Nielsen-Olesen model. One notices from Figs. \ref{figv2}, \ref{figv4} and \ref{figv5} that the dependence of the vortex on $\lambda$ is much more significant for $\lambda\in[0,1]$. Nevertheless, the results show that the parameter $\lambda$ provides the possibility to construct vortices with profile different from the Nielsen-Olesen one. Since the function $f(\chi)$ that appear in \eqref{lvortex} is directly related to the magnetic permeability \cite{V1} of the model, it is the change in the magnetic permeability that responds for the modification of the vortex configuration. We also notice that the total energy of the field configurations, which appears just above Eq. \eqref{fochi}, does not depend on $\lambda$ and $m$, and this leads us to the possibility to control the profile of the solution without changing its total energy. On the other hand, the total energy depends on $\alpha$, which controls the shape of the rings and is related to the additional $Z_2$ symmetry.

We can consider another possibility, changing the function $f(\chi)$ to the new form $f(\chi)=1/J_1^2(\gamma\chi) $, $\gamma\in \mathbb{R}$, with $J_1$ being the Bessel function of first kind. We call this the Bessel case, which is motivated by Ref. \cite{RVprl} that deals with a Bessel lattice. We shall further comment on this in the next Section, but here the first order equations \eqref{fovmult} change to, taking $r_0=1$,
\begin{equation}
g^\prime=\frac{ag}{r}, \;\;\;\;\;-\frac{a^\prime}{r} = J_1^2\!\left(\gamma\, \frac{r^{2\alpha}-1}{r^{2\alpha}+1}\right)\!\left(1-g^2\right).
\end{equation}
These equations are solved numerically, and the magnetic field is now depicted in the plane in Fig. \ref{figv6}. We compare these results with the previous ones to see that the magnetic permeability can be modulated to control how the magnetic field spread around the core of the localized vortex solution.

It is important to notice that although the magnetic field may be spread around the center of the vortex in different ways, its energy, stability, topological charge and magnetic flux remain the same.

\subsubsection{Cubic and quintic nonlinearities}
 
\begin{figure}[t!]
		\centering
		\includegraphics[width=4.2cm]{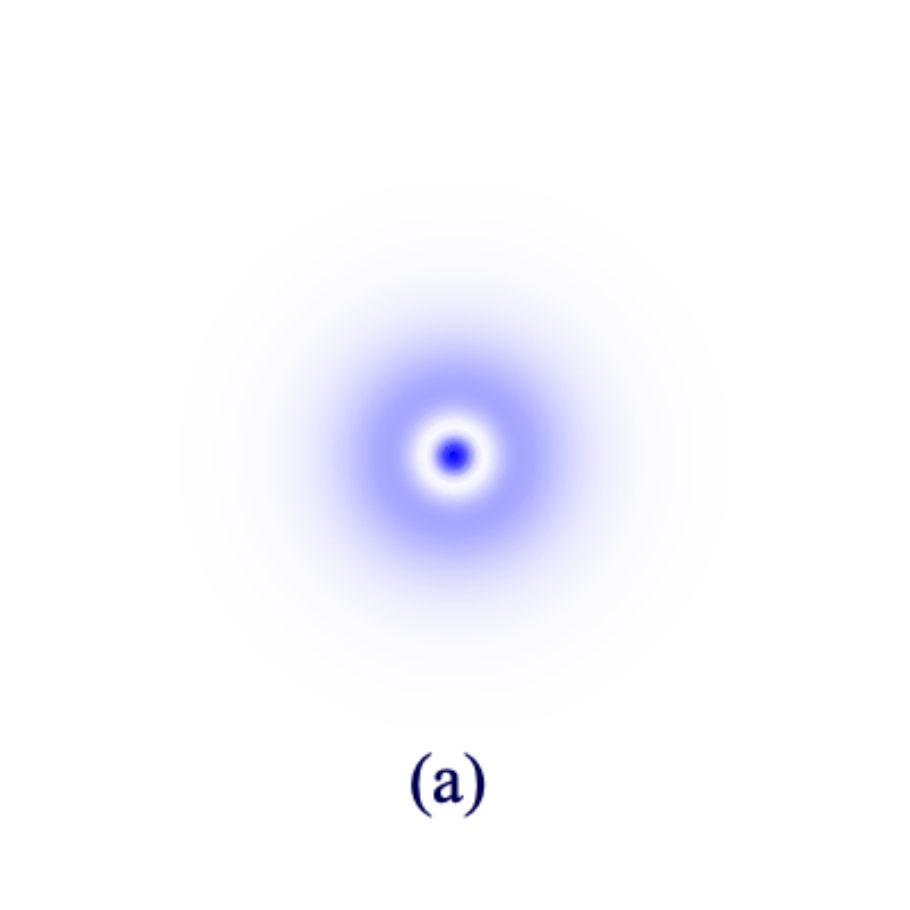}
		\includegraphics[width=4.2cm]{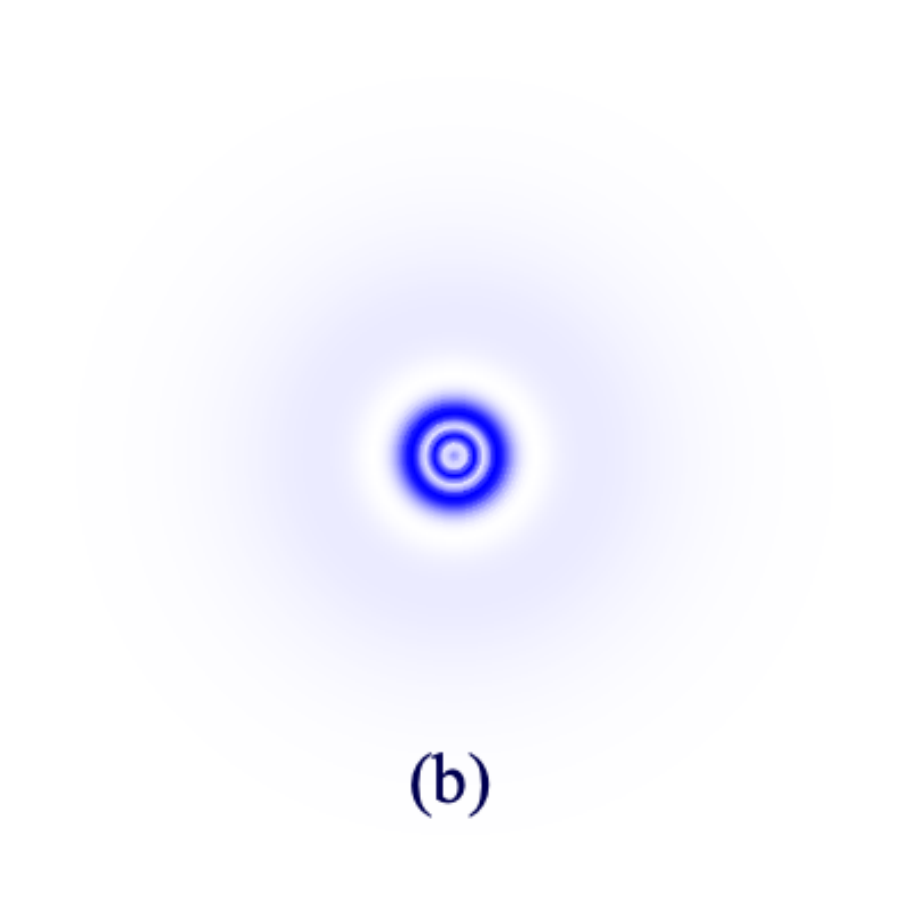}
		\caption{The first model. The magnetic field is depicted in the plane for the case of cubic nonlinearity in the Bessel case, with $\alpha=1$ and $\gamma=1.5$ (a) and $5$ (b).}
		\label{figv6}
		\end{figure}

\begin{figure}[t!]
		\centering
		\includegraphics[width=4.2cm]{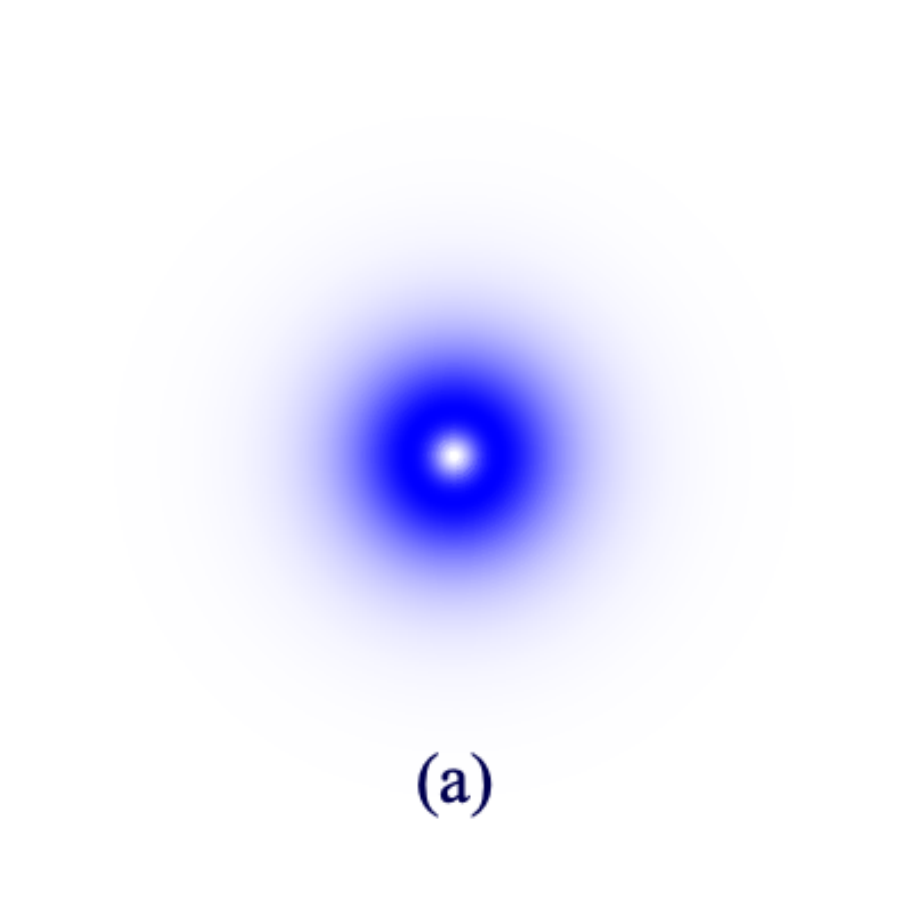}
		\includegraphics[width=4.2cm]{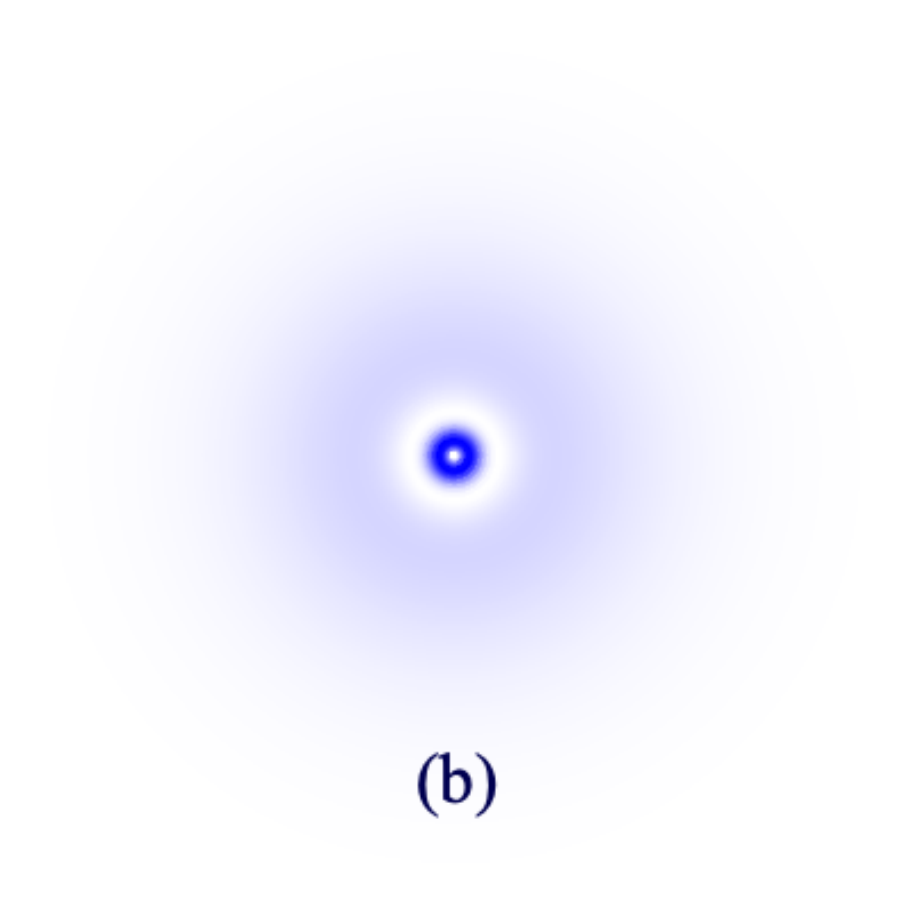}
		\caption{The first model. The magnetic field is depicted in the plane for the case of cubic and quintic nonlinearity in the Bessel case, with $\alpha=1$ and $\gamma=1.5$ (a) and $5$ (b).}
		\label{figv7}
		\end{figure}

We can study another type of potential for the $\chi$ field. We do this changing \eqref{chi4} to
\be 
W_\chi =\alpha \chi(1-\chi^2)
\ee
In this case, the minima are also at $\pm1$, but now there is another minimum at $\chi=0$. The equation of motion of the $\chi$ field engenders cubic and quintic nonlinearities and the solution changes from \eqref{solvchi} to the new one 
\be 
\chi(r)=\frac{r^{\alpha}}{\sqrt{r_0^{2\alpha}+r^{2\alpha}}}.
\ee
Here we also take $r_0=1$. In this case, $\chi(r)$ is zero at the origin and so different from the previous case. We then investigate how the change from the cubic case displayed before to the above cubic and quintic situation reflects in the profile of the vortex configuration. To investigate this we consider the same Bessel case, $f(\chi)=1/J_1^2(\gamma\chi)$, as we did before. In this case the first order equation \eqref{fovmult} changes to
\be 
g^\prime=\frac{ag}{r}, \;\;\;\;\;-\frac{a^\prime}{r} = J_1^2\!\left(\frac{\gamma \,r^{\alpha}}{\sqrt{1+r^{2\alpha}}}\right)\!\left(1-g^2\right),
\ee
and the magnetic field is depicted in the plane in Fig. \ref{figv7}. We can compare the results of Figs. \ref{figv6} and \ref{figv7} to see that the nonlinearity associated to the $\chi$ field is also important to control the magnetic field around the center of the vortex solution.

We summarise the above results noticing that both the magnetic permeability and the nonlinearity associated to the $\chi$ field are important to modify the solution and control the distribution of magnetic field around its center, changing the standard solution into a multilayered structure.       

\subsection{Second model, with $U(1)\times U(1)$ symmetry}

Let us now consider another model. If instead of the $Z_2$ symmetry we use another local $U(1)$ symmetry, the model changes to accommodate extra fields, another complex scalar field $\chi$ and another gauge field $\A_\mu$. In this case, the Lagrange density becomes
\be\label{lu1u1}
\begin{aligned}
	\LL &= - \frac{1}{4}f(|\chi|)F_{\mu\nu}F^{\mu\nu} - \frac{1}{4}\F_{\mu\nu}\F^{\mu\nu} \\
	    &\hspace{4mm} + |D_\mu\vphi|^2 + |\D_\mu\chi|^2 - V(|\vphi|,|\chi|),
\end{aligned}
\ee
where we use the notation $\F_{\mu\nu}=\partial_\mu \A_\nu-\partial_\nu \A_\mu$ and $\D_\mu=\partial_{\mu}+iq\A_{\mu}$. Since we aim to deal with vortex configurations, we take static fields. Moreover, we consider the ansatz $A_0=\A_0=0$, with $\vphi$ and $\vec{A}$ given as in Eq.~\eqref{ansatz1} and the other fields as
\begin{equation}
\chi = h(r)e^{ik\theta}	,\;\;\;\;\;\vec{\A} = -{\frac{\hat{\theta}}{qr}\,(c(r)-k)},
\end{equation}
with $k$ being another nonvanishing integer. The boundary conditions now are: $h(0)=0$, $h(\infty)=w$, $c(0)=k$ and $c(\infty)=0$. We see that one is doubling the degrees of freedom used to describe the standard vortex solution, and the flux associated to the additional gauge field $\vec{\A}$ is also quantized:
$\Phi_{\!\vec{\A}} = (2\pi/q) k $. The equations of motion are, taking $n=k=1$,
\bes\label{eomu1u1}
\begin{align}
&\frac{1}{r} \left(r g^\prime\right)^\prime -\frac{a^2g}{r^2} - \frac12  V_g = 0, \\ 
&\frac{1}{r} \left(r h^\prime\right)^\prime\! -\!\frac{c^2h}{r^2}\! -\! \frac12 \left(\! f_h\frac{{a^\prime}^2}{2r^2}\! +\! V_h\!\right)\! = 0, \\
&r\left(f\frac{a^\prime}{r} \right)^\prime - 2ag^2 = 0, \\
&r\left(\frac{c^\prime}{qr} \right)^\prime - 2qch^2 = 0.
\end{align}
\ees
The energy density of the static fields is
\be
\rho=f(h) \frac{{a^\prime}^2}{2r^2} + {g^\prime}^2 +\frac{a^2g^2}{r^2} + \frac{{c^\prime}^2}{2q^2r^2} + {h^\prime}^2 + \frac{c^2h^2}{r^2} + V(g,h).
\ee

		\begin{figure}[t!]
		\centering
		\includegraphics[width=4.2cm]{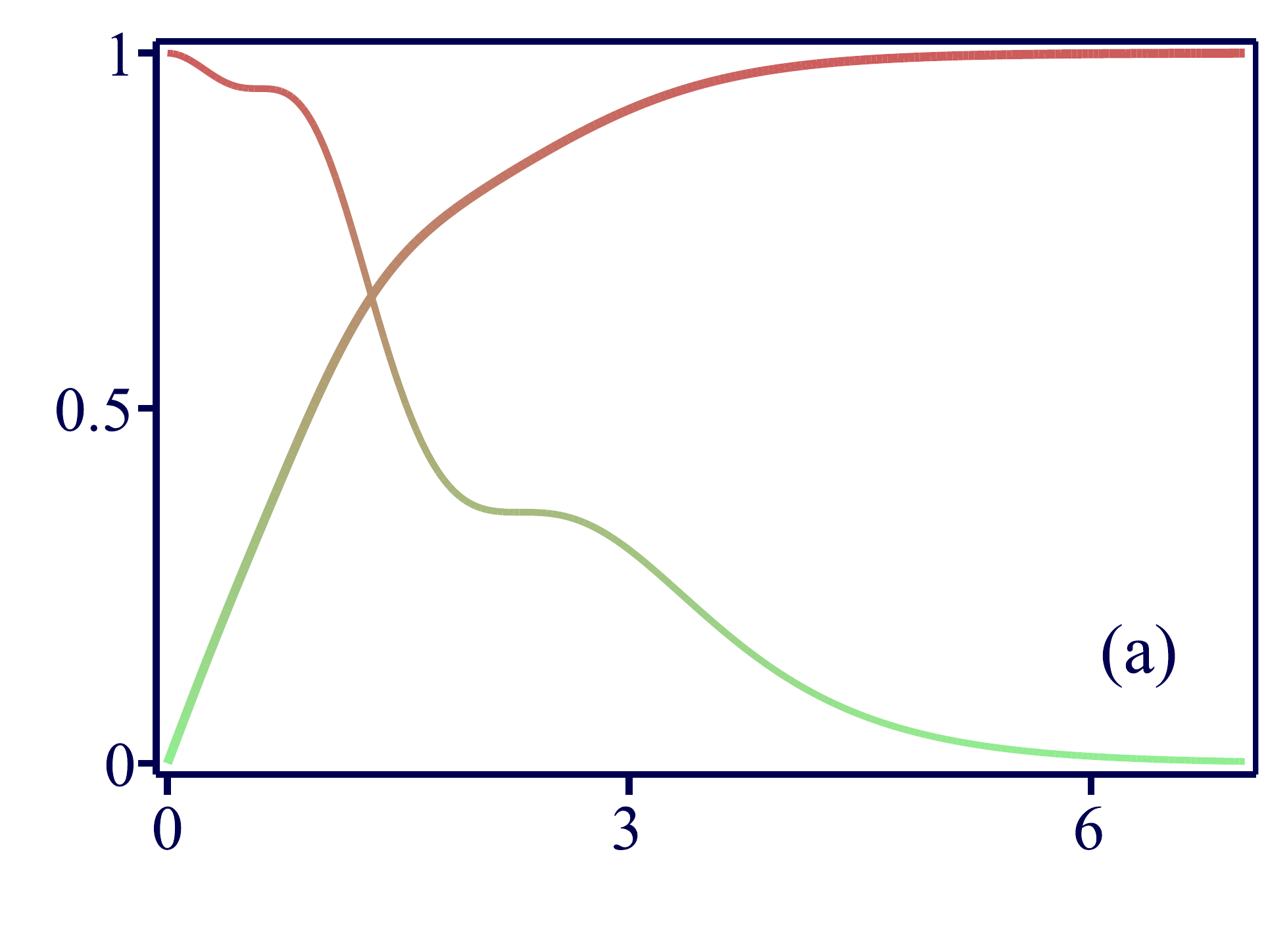}
		\includegraphics[width=4.2cm]{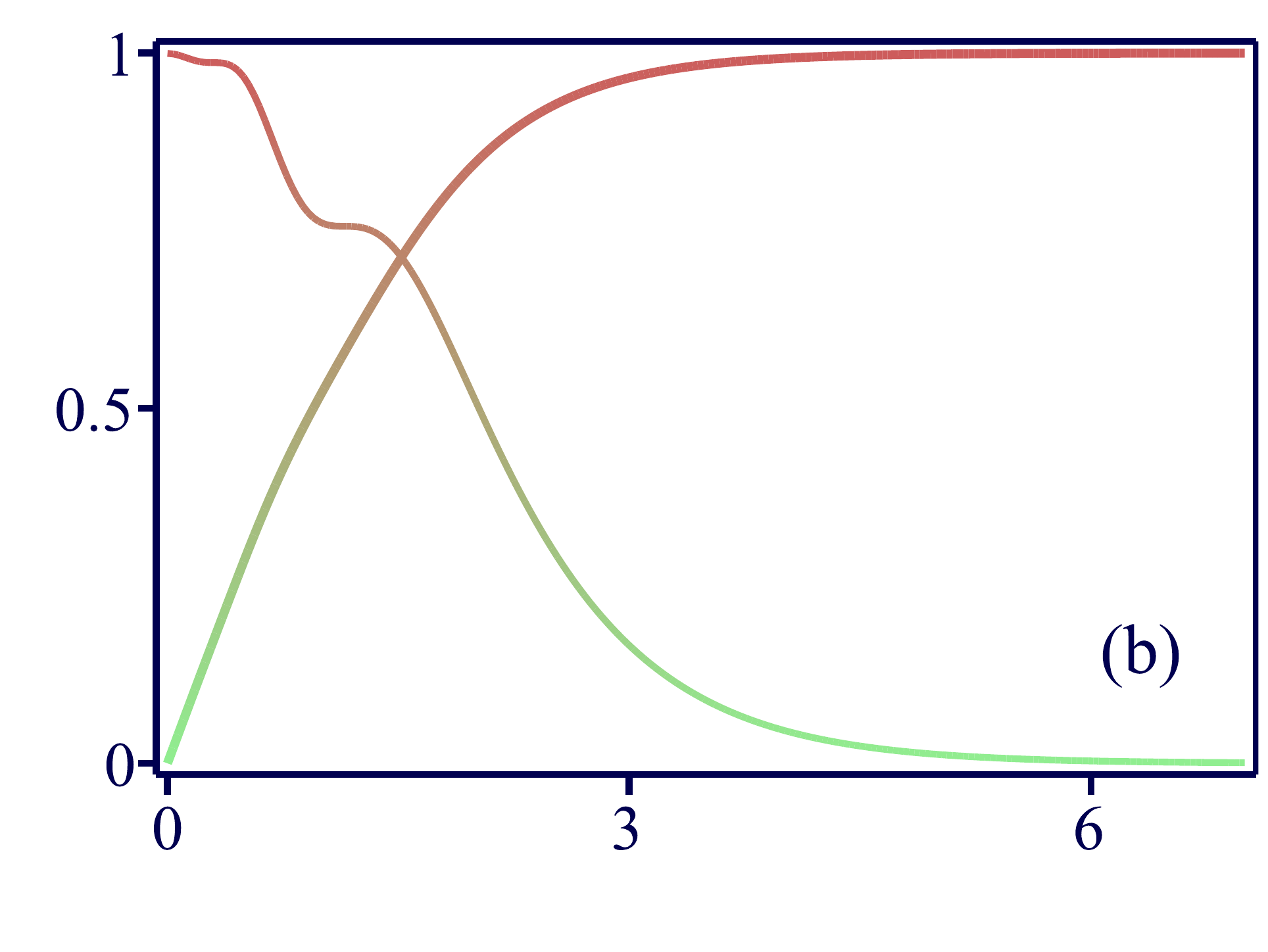}
		\includegraphics[width=4.2cm]{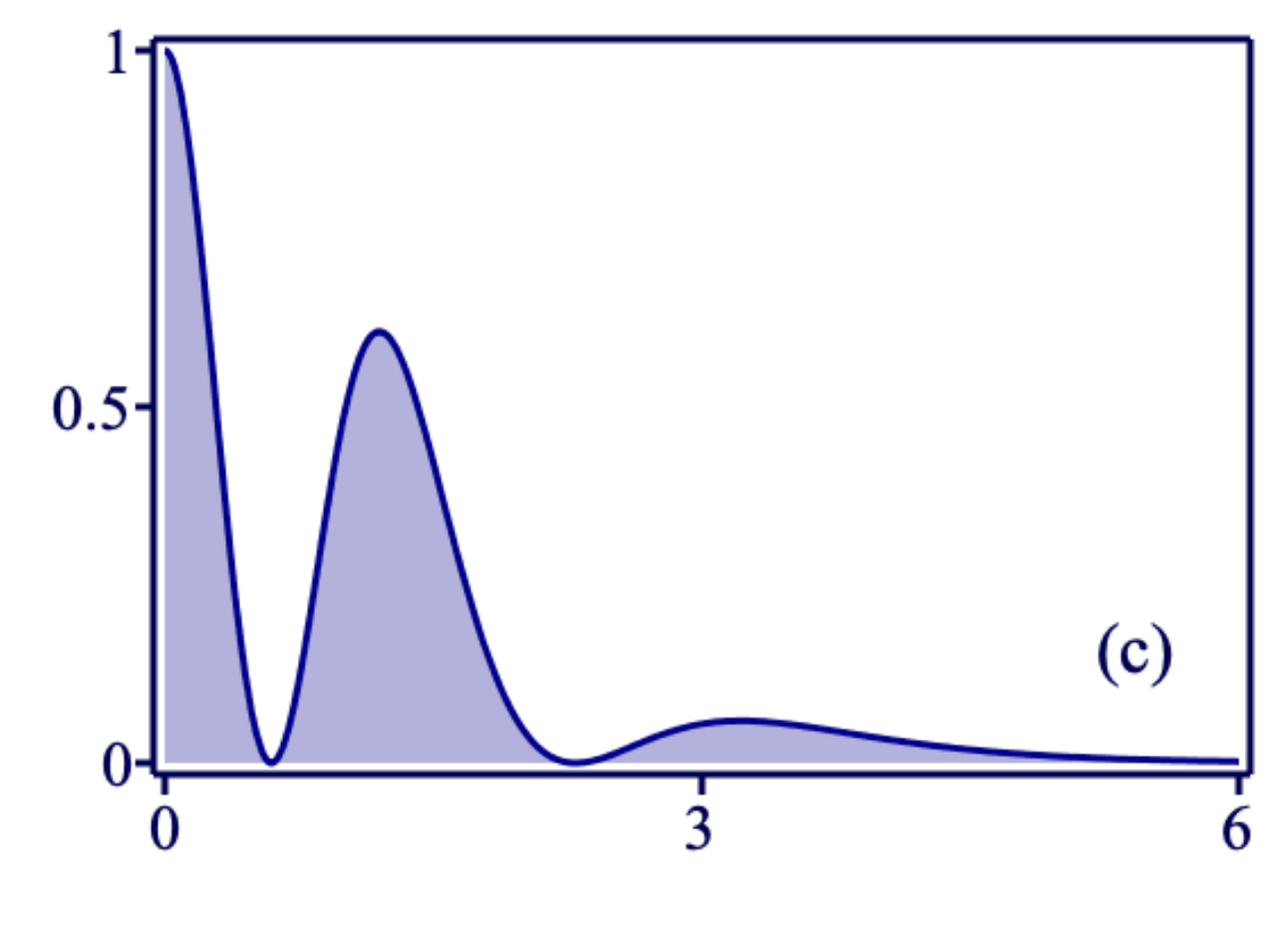}
		\includegraphics[width=4.2cm]{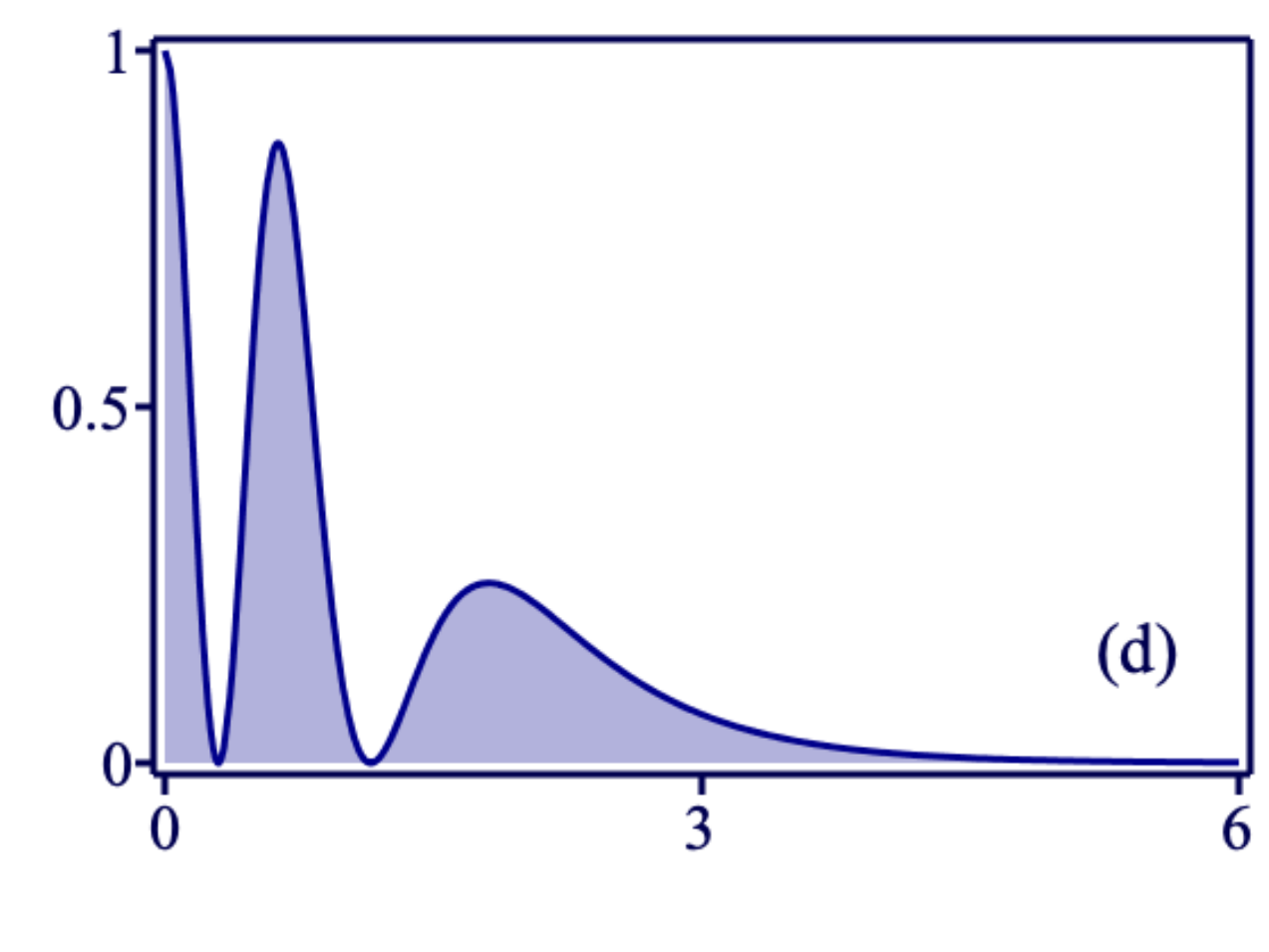}
		\includegraphics[width=4.2cm]{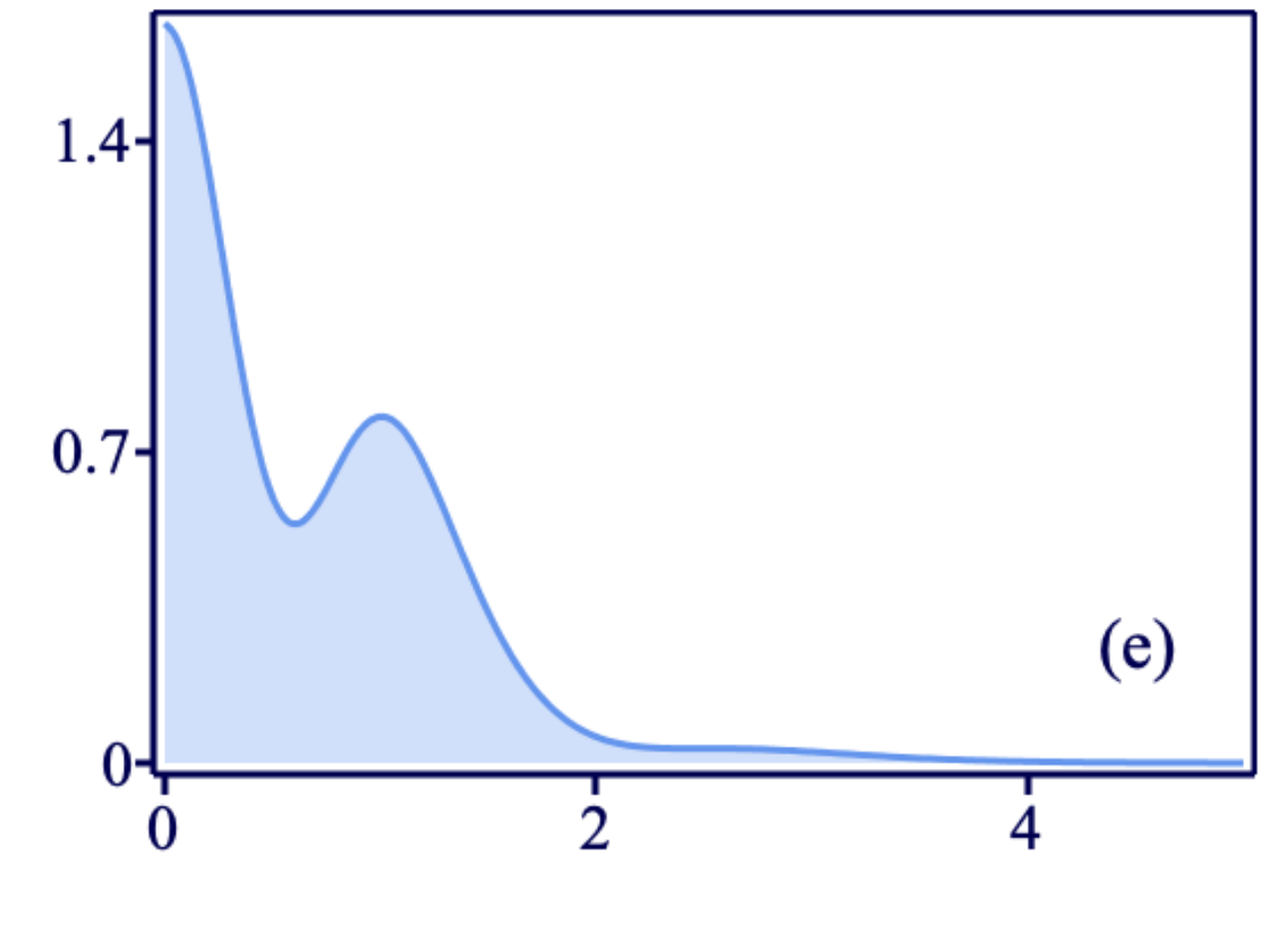}
		\includegraphics[width=4.2cm]{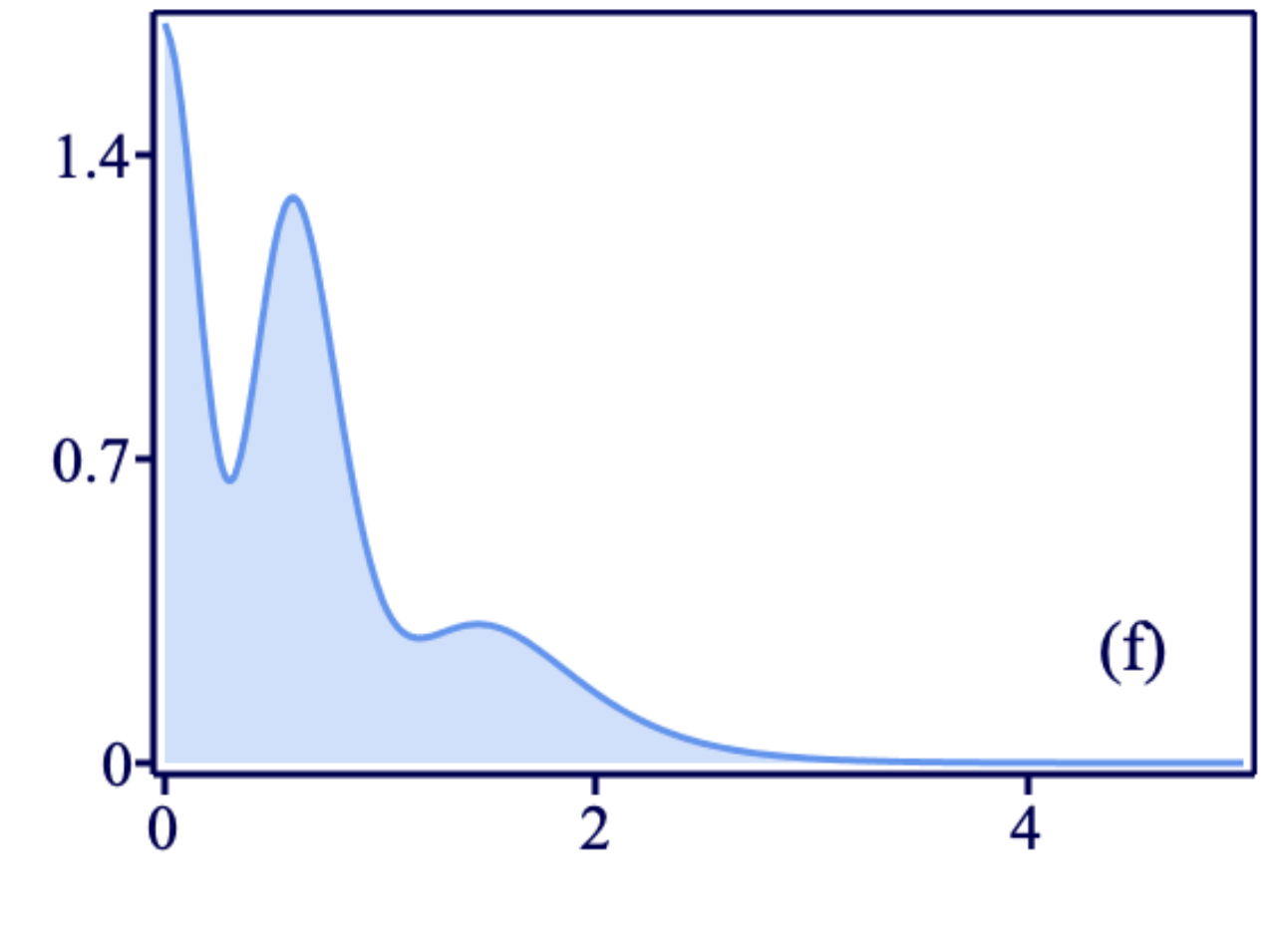}
		\caption{The second model. The vortex solutions $a(r)$ (descending line) and $g(r)$ (ascending line) in the panels (a) and (b), their magnetic field $B=-a^\prime/r$ ((c) and (d)) and their energy density $\rho_{2}(r)$ ((e) and (f)). They are depicted for $n=1$, $k=1$, $\lambda=0$, $m=2$ and $q=0.5$ (left) and $1$ (right).}
		\label{figv8}
		\end{figure}

		\begin{figure}[t!]
		\centering
		\includegraphics[width=4.2cm]{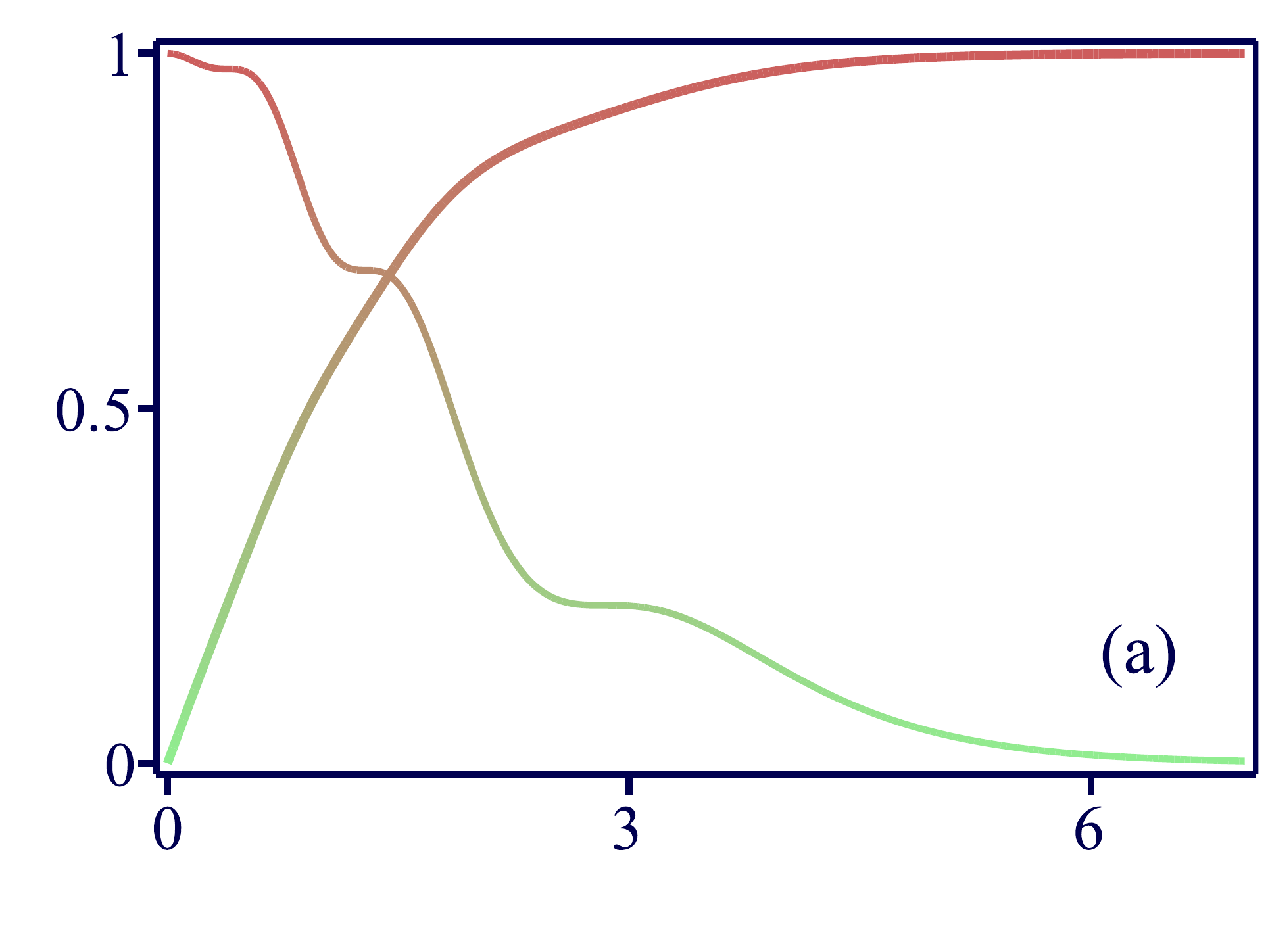}
		\includegraphics[width=4.2cm]{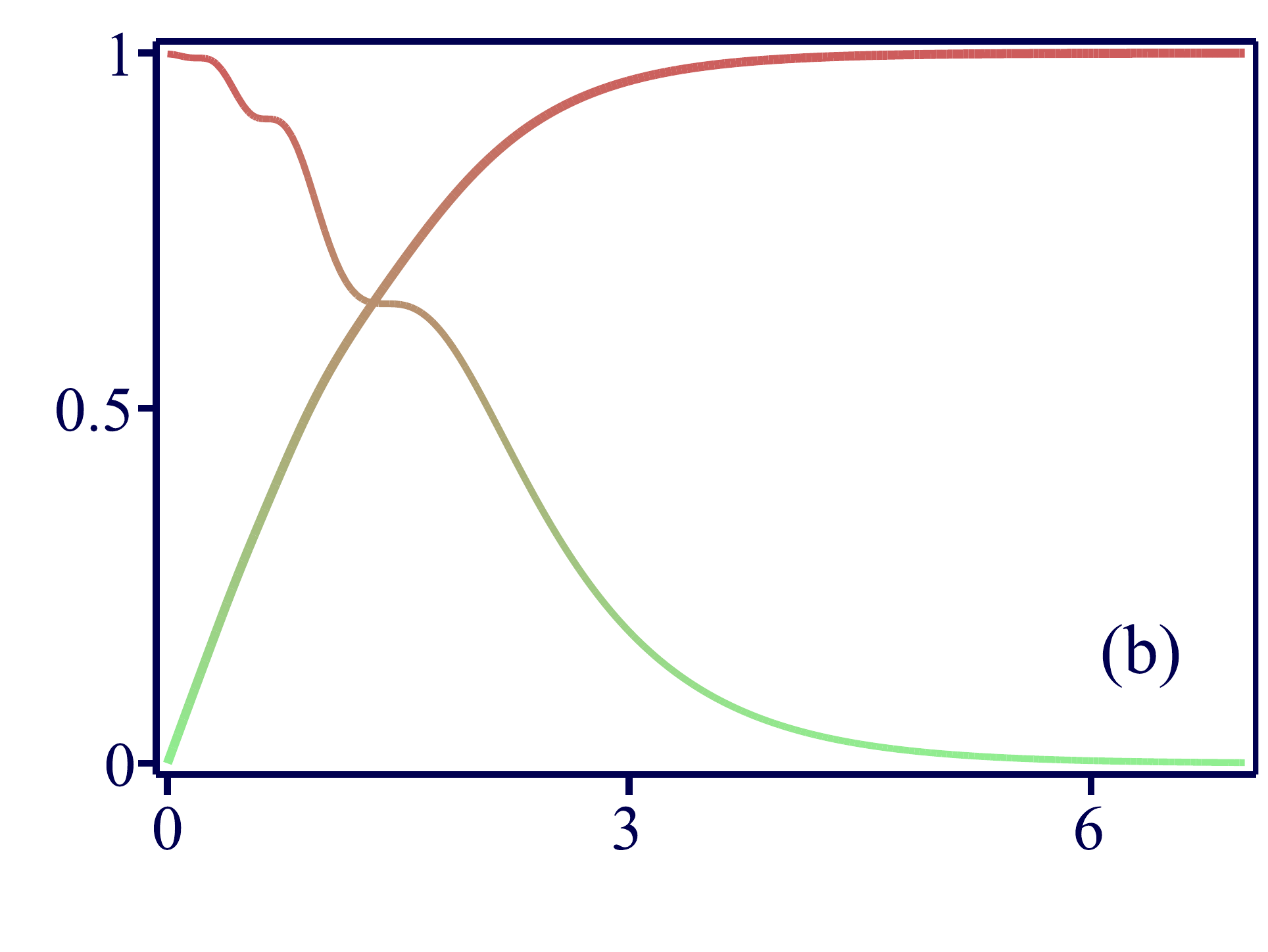}
		\includegraphics[width=4.2cm]{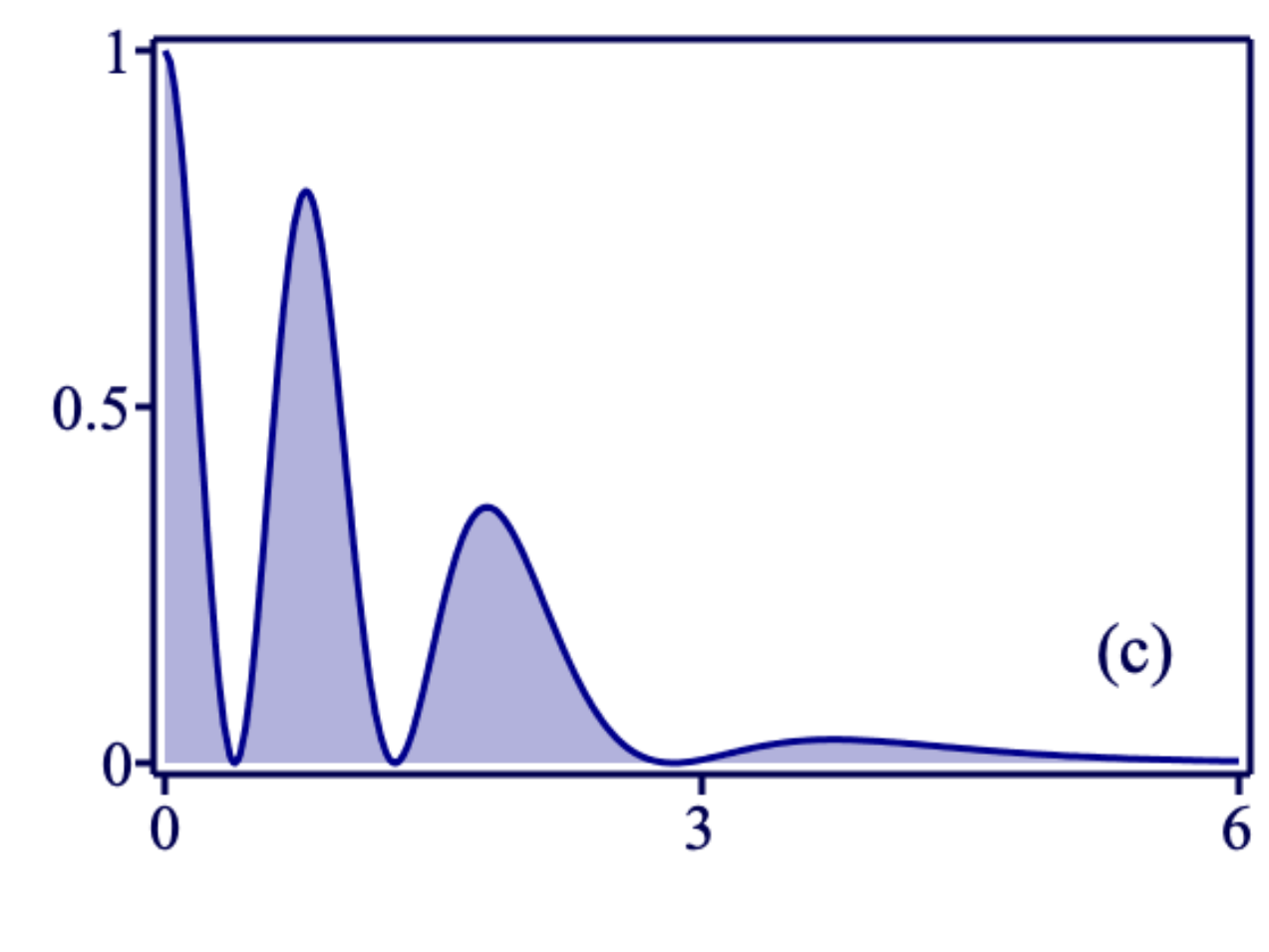}
		\includegraphics[width=4.2cm]{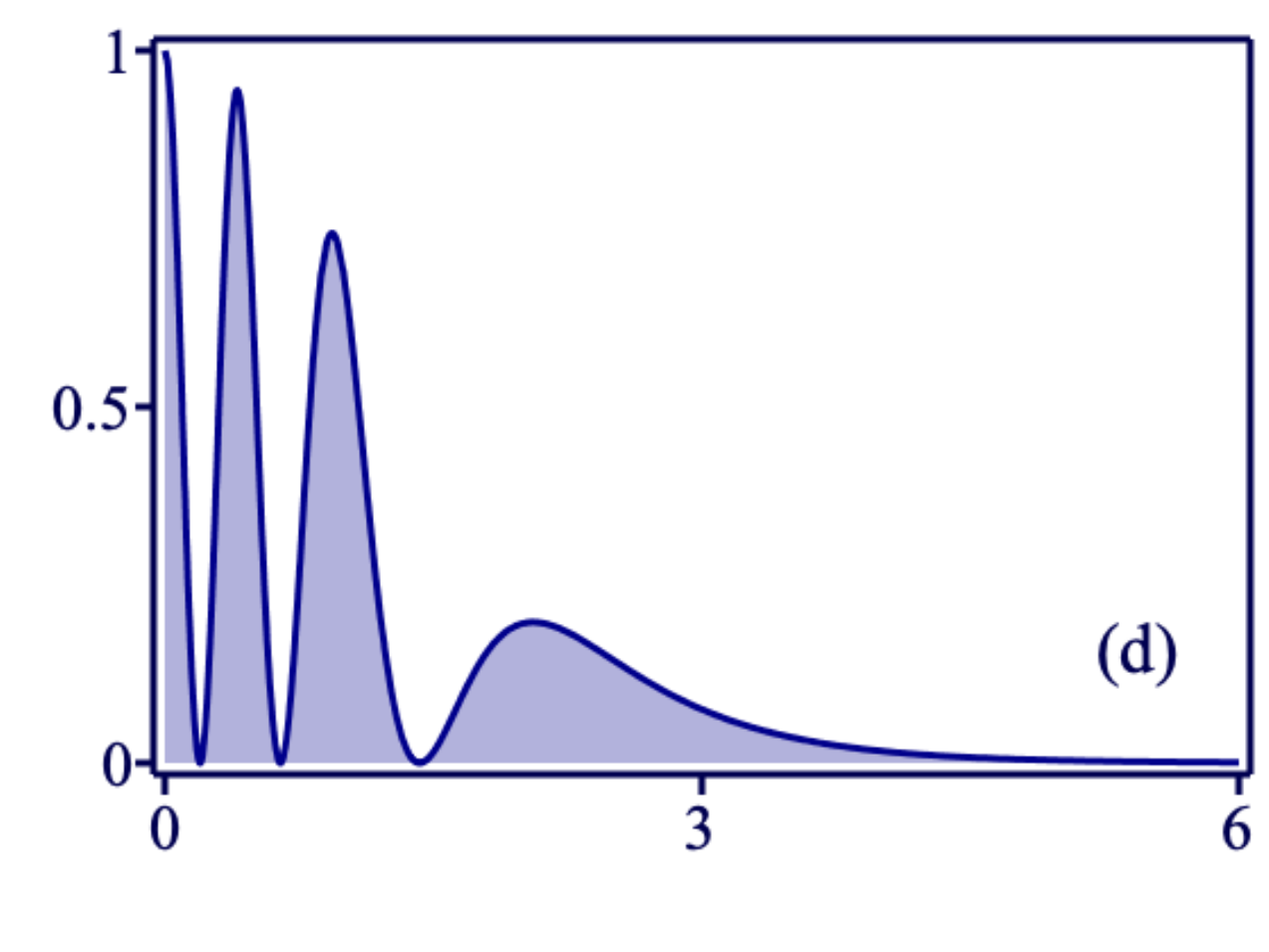}
		\includegraphics[width=4.2cm]{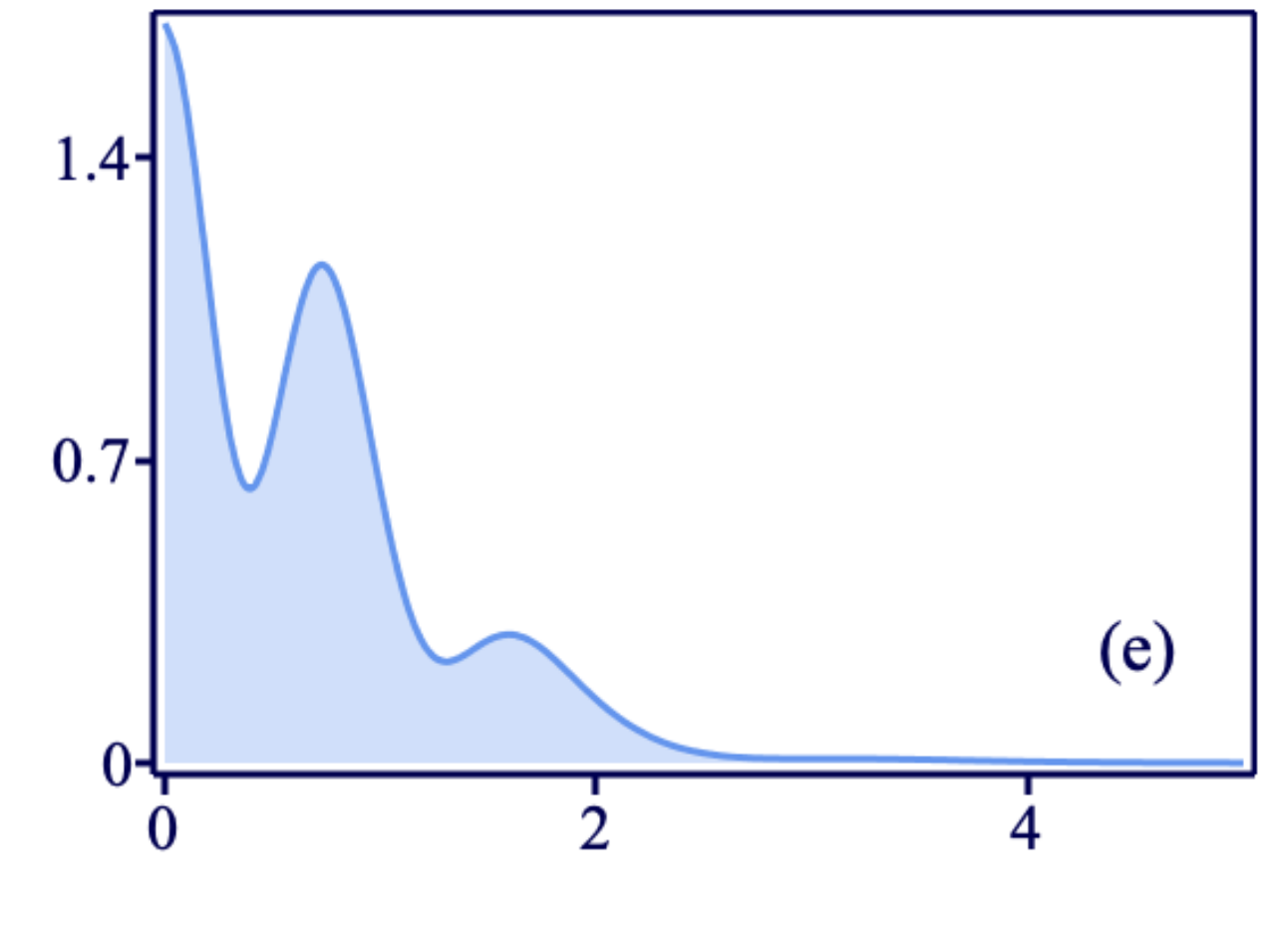}
		\includegraphics[width=4.2cm]{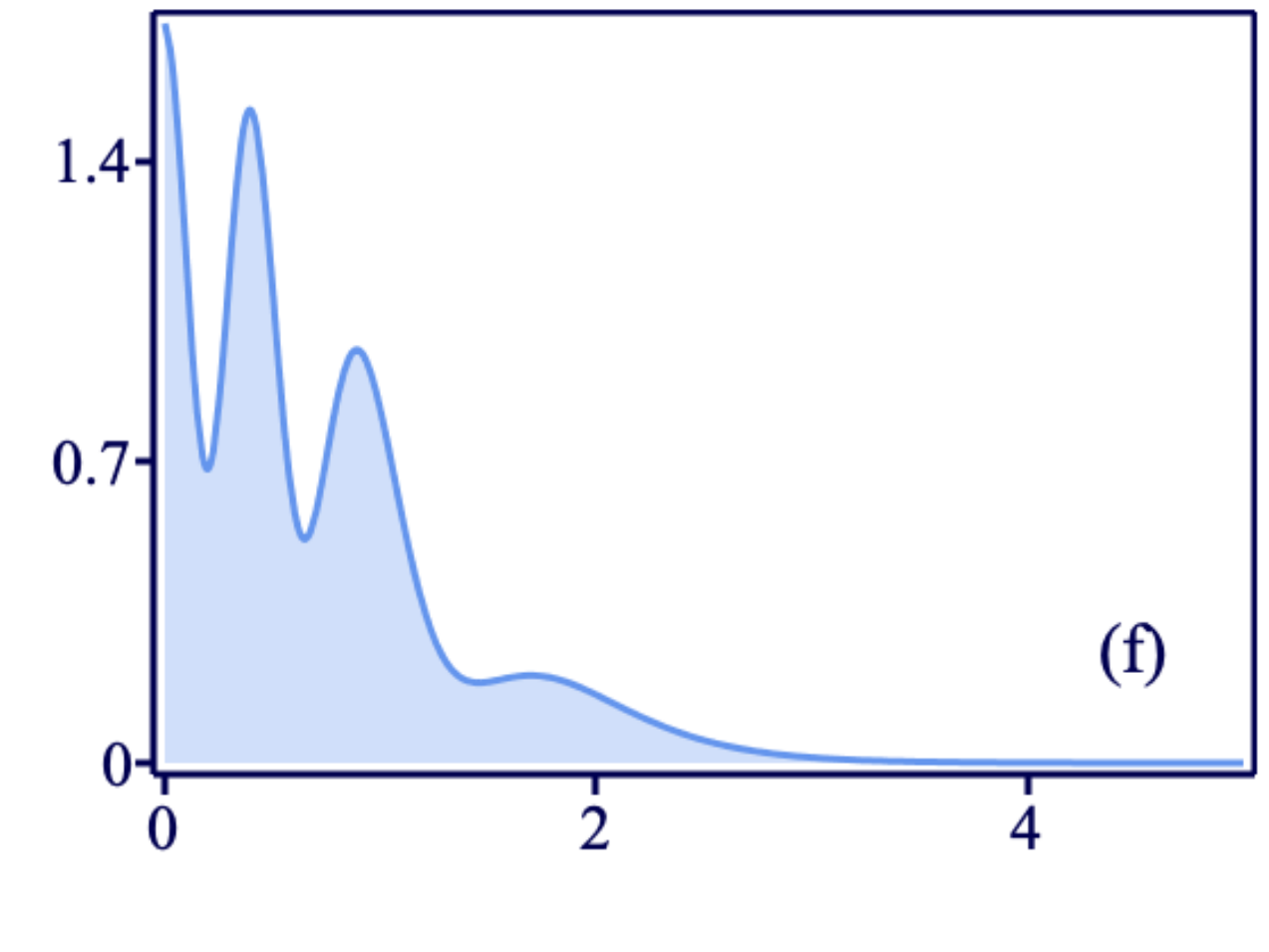}
		\caption{The second model. The vortex solutions $a(r)$ (descending line) and $g(r)$ (ascending line) in the panels (a) and (b), the magnetic field $B(r)=-a^\prime/r$ ((c) and (d)) and the energy density $\rho_{2}(r)$ ((e) and (f)). They are depicted for $n=1$, $k=1$, $\lambda=0$, $m=3$ and $q=0.5$ (left) and $1$ (right).}
		\label{figv9}
		\end{figure}

To find a first order framework, one follows \cite{ahep} to show that the potential
\be\label{pote}
V(|\vphi|,|\chi|) =  \frac{1}{2} \frac{\left(1-|\vphi|^2\right)^2}{f(|\chi|)}  + \frac{q^2}{2} \left(w^2-|\chi|^2\right)^2,
\ee
turns the energy minimized to $2\pi\left(1+ w^2\right)$ if the following first order equations are satisfied
\be\label{fovisible}
g^\prime = \pm\frac{ag}{r},\;\;\;\;\;\;\;\;
-\frac{a^\prime}{r} = \pm \frac{\left(1-g^2\right) }{f(h)},
\ee
and
\be\label{fohidden}
h^\prime = \pm\frac{ch}{r},\;\;\;\;\;\;\;\;
-\frac{c^\prime}{qr} = \pm q \left(w^2-h^2\right).
\ee
The upper and lower signs are related by the change $a\to-a$ and $c\to-c$, so we only deal with positive signs. Notice the first order equations \eqref{fohidden} depend exclusively on $c(r)$ and $h(r)$. So, we solve them independently, and use the solution to feed the magnetic permeability controlled by the function $f(h(r))$ in Eqs.~\eqref{fovisible} and then find the solutions $a(r)$ and $g(r)$. For simplicity, from now on we consider $w=1$. The first order equations \eqref{fovisible} and \eqref{fohidden} allows us to write the energy density \eqref{rhoans} in the form $\rho = \rho_{1} + \rho_{2}$, where
\be\label{rhoh}
\rho_{1} =  \frac{{c^\prime}^2}{q^2r^2} + 2{h^\prime}^2,\;\;\;\;\;
\rho_{2} = f(h) \frac{{a^\prime}^2}{r^2} + 2{g^\prime}^2.
\ee
We notice that the localized structures determined by Eqs.~\eqref{fohidden} are the Nielsen-Olesen vortex solutions for $q=1$, which we call standard solutions and identify with the subscript ${st}$; thus, $c(r)=c_{st}(qr)$ and $h(r)=h_{st}(qr)$. This shows that the parameter $q$ shrinks or expands the standard vortex solutions, and since these solutions are well-known, with its corresponding energy density $\rho_1$ in Eq.~\eqref{rhoh} being a hump, we do not depict them here. 

To investigate how the aforementioned solutions modify the other vortex structure, we also consider
$f(|\chi|) = (1+\lambda^2)/(\lambda^2+\cos^2(2\pi m|\chi|))$, $m\in\mathbb{N}$ and $\lambda \in \mathbb{R}$. In the case for $\lambda=0$ the first order equations \eqref{fovisible} become
\be\label{fovu1u1}
g^\prime = \frac{ag}{r},\;\;\;\;\;
-\frac{a^\prime}{r} = \cos^2\left(2\pi m\, h_{st}(qr)\right)\left(1-g^2\right).
\ee
We use numerical procedures and depict in Figs.~\ref{figv8} and \ref{figv9}, the vortex solutions, magnetic field, $B=-a^\prime/r$, and energy density $\rho_2(r)$ for some specific values of the parameters.

\begin{figure}[t!]
		\centering
		\includegraphics[width=4.2cm]{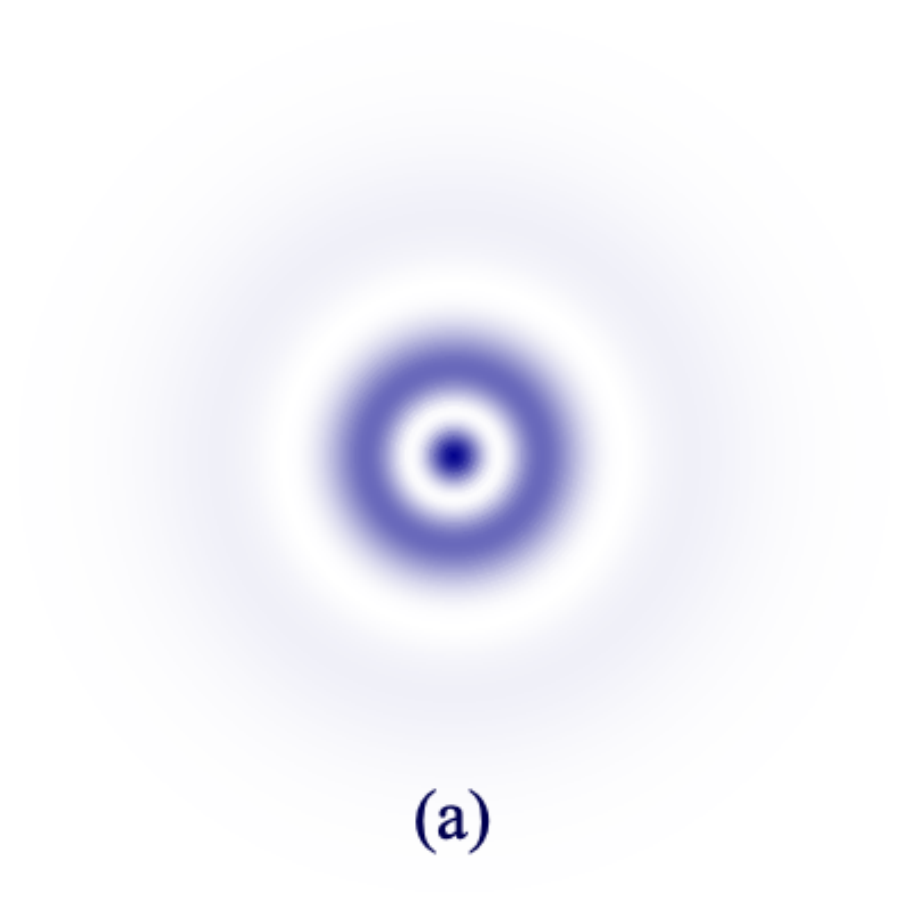}
		\includegraphics[width=4.2cm]{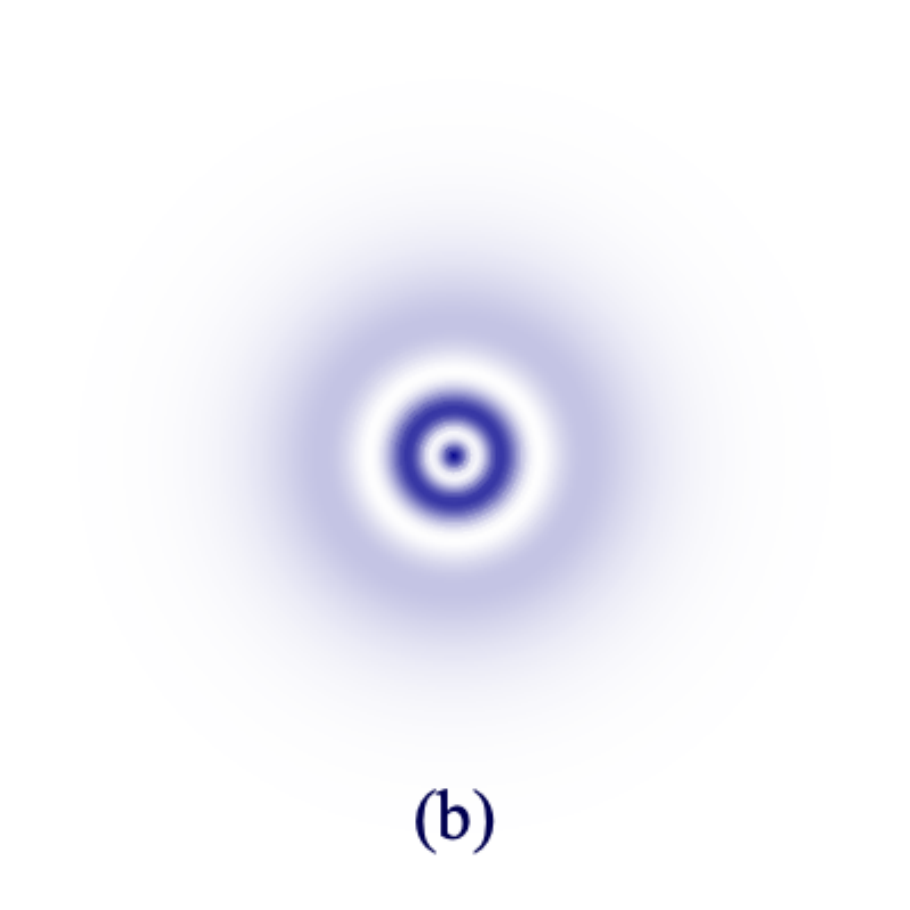}
		\includegraphics[width=4.2cm]{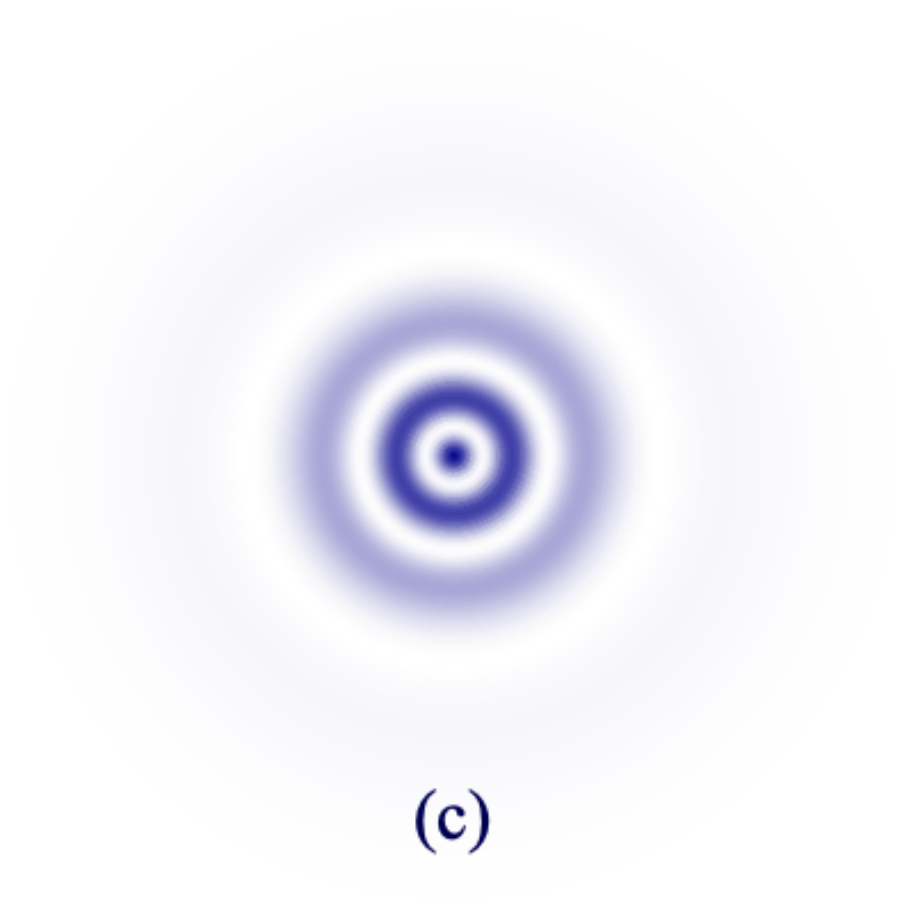}
		\includegraphics[width=4.2cm]{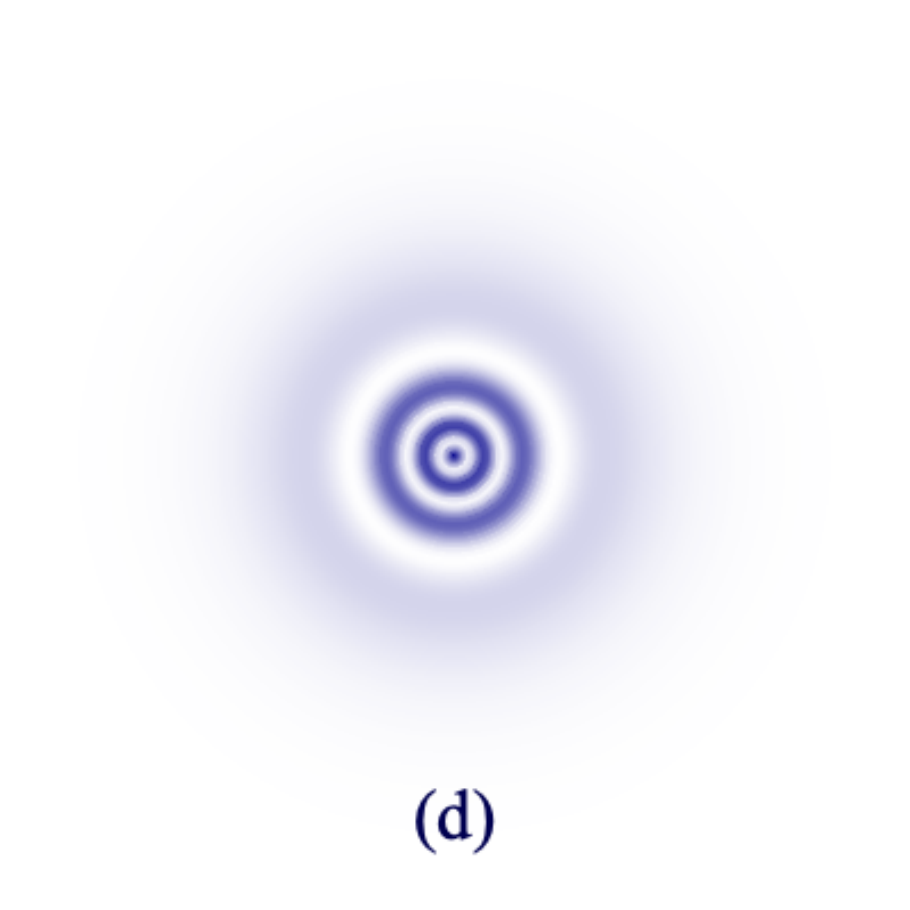}
		\caption{The second model. The magnetic field associated to the vortex is depicted in the plane for $n=1$, $k=1$, $\lambda=0$, $q=0.5$ and $m=2$ (a), $q=1$ and $m=2$ (b), $q=0.5$ and $m=3$ (c), and $q=1$ and $m=3$ (d).}
		\label{figv10}
		\end{figure}

\begin{figure}[t!]
		\centering
		\includegraphics[width=4.2cm]{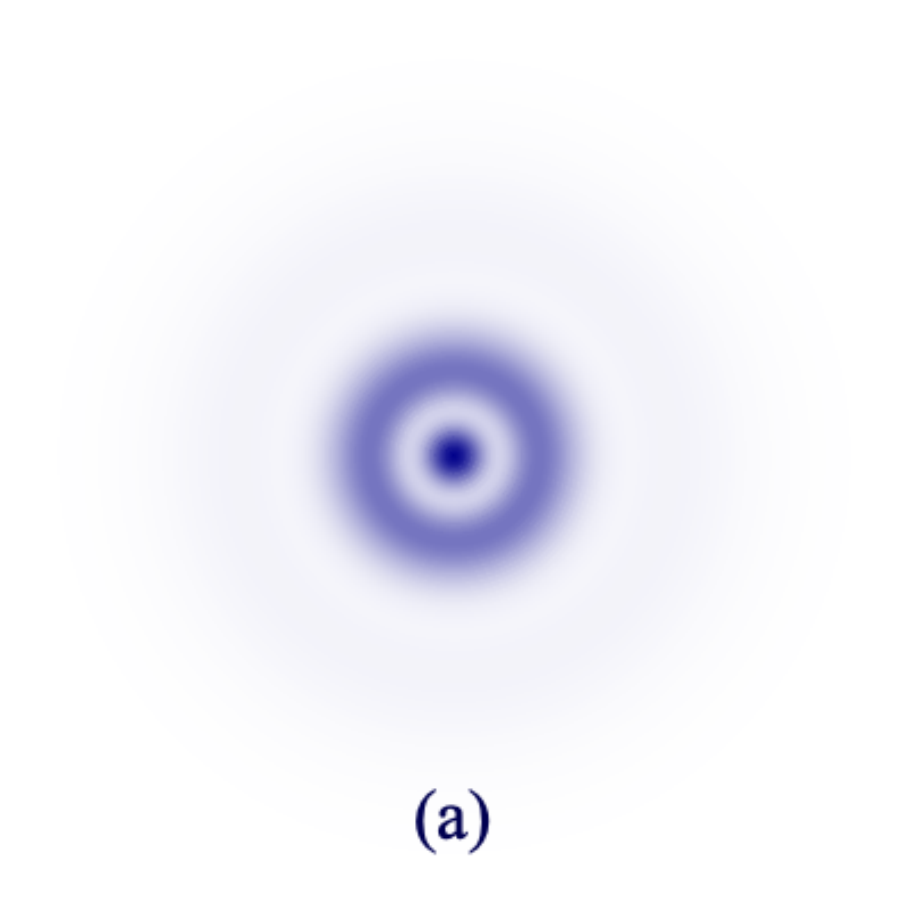}
		\includegraphics[width=4.2cm]{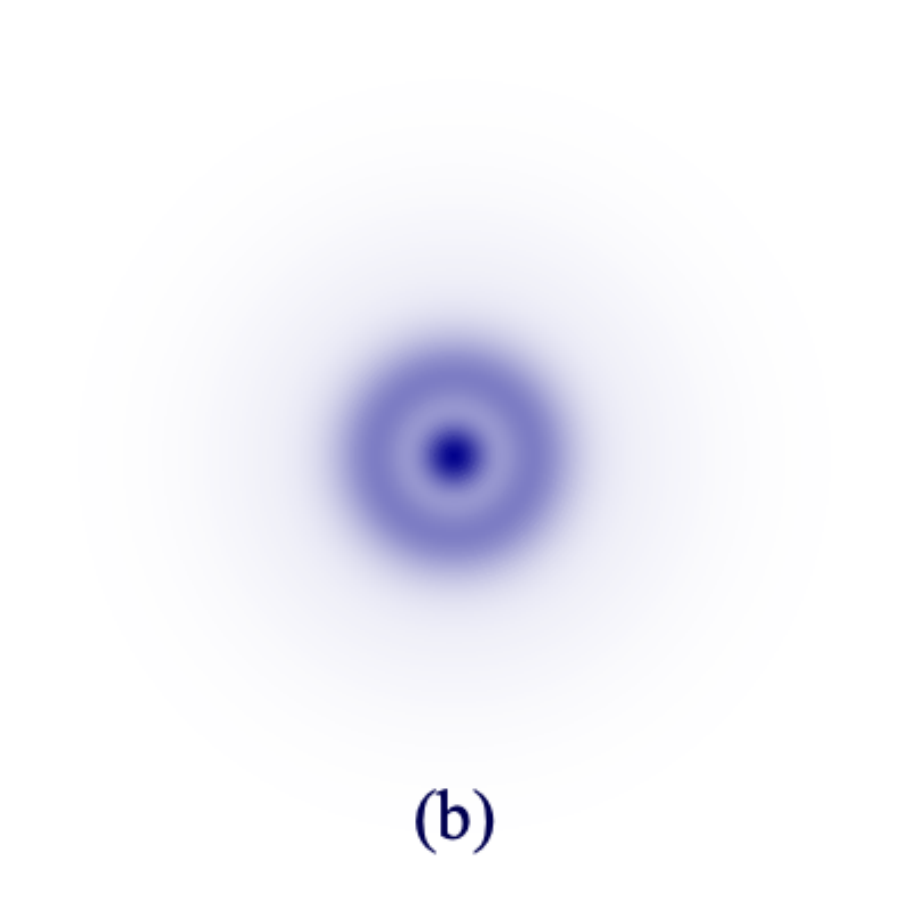}
		\includegraphics[width=4.2cm]{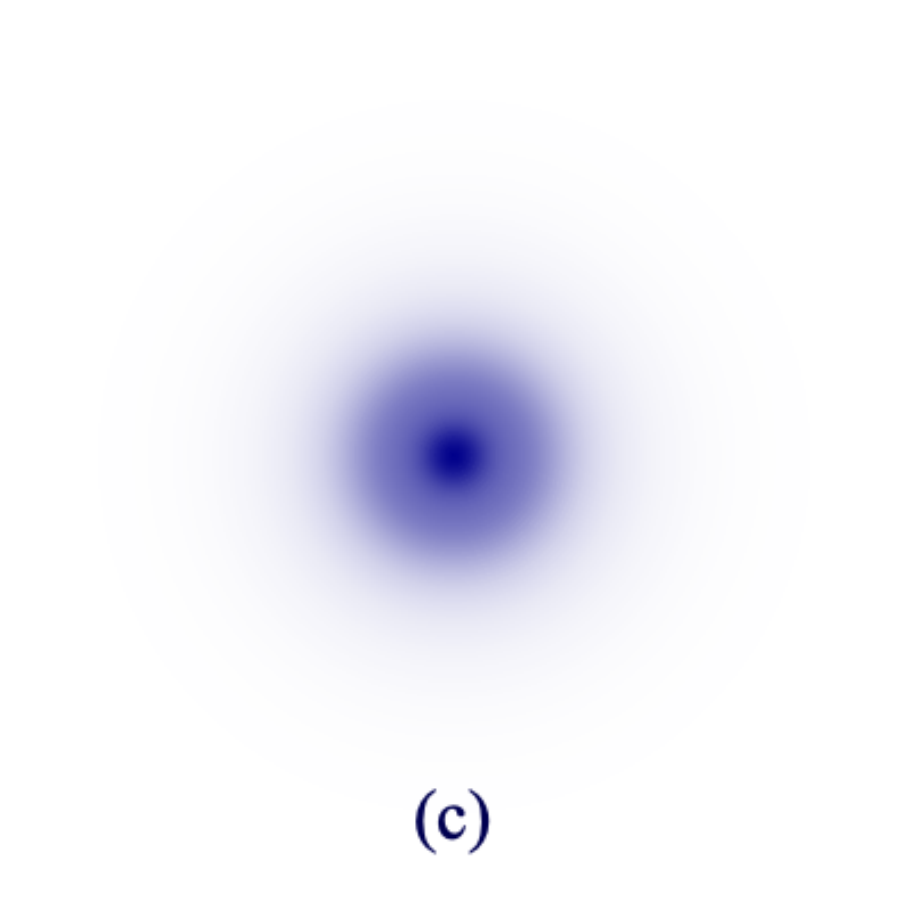}
		\includegraphics[width=4.2cm]{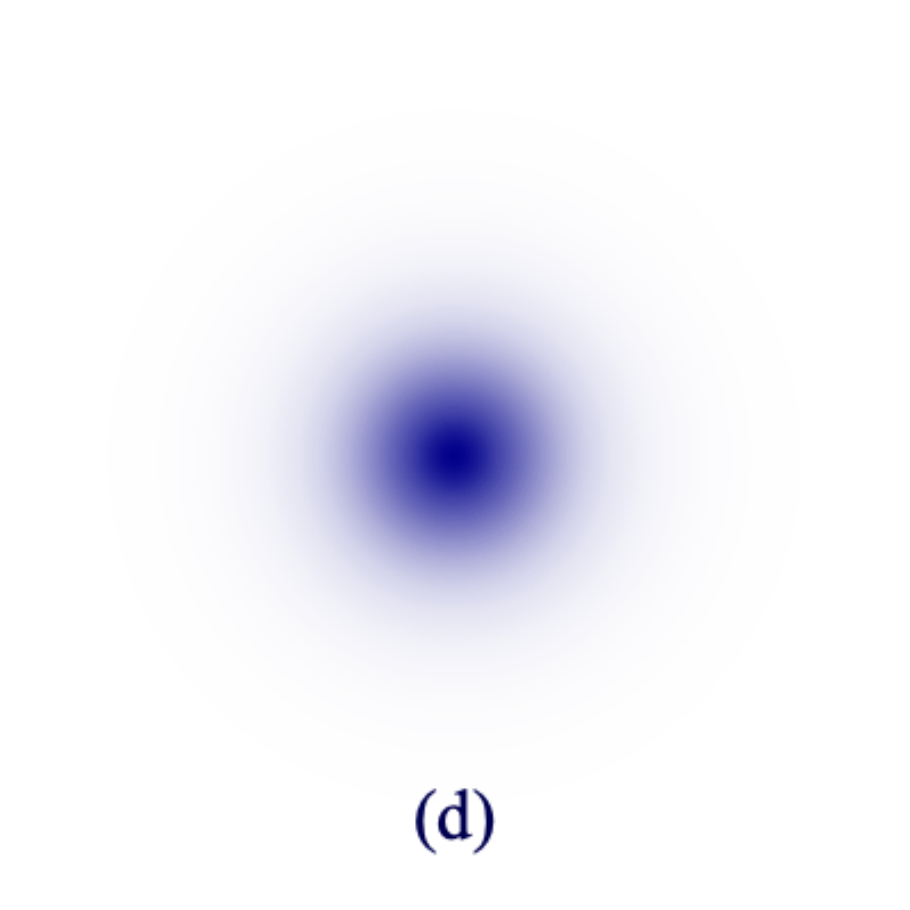}
		\caption{The second model. The magnetic field associated to the vortex is depicted in the plane for $n=1$, $k=1$, $q=0.5$, $m=2$, and $\lambda=0.5$ (a), $1$ (b), $2$ (c), and $4$ (d).}
		\label{figv11}
		\end{figure}

As in the previous model, here the magnetic field also presents internal structures that are controlled by the parameters $q$ and $m$. In Fig.~\ref{figv10}, we depict the magnetic field in the plane, displaying a central disk and $m$ rings around it, which are also controlled by $q$. The role played by $\lambda$ is similar to the previous case; for this reason, in Fig. \ref{figv11} we only depict the magnetic field in the plane for $q=0.5$, $m=2$, and $\lambda=0.5, 1, 2$ and $4$. The results show that the ringlike structures are much more evident as $\lambda$ approaches zero. We also notice that the total energy which appears below the potential given by Eq. \eqref{pote} does not depend on $q$, which is provided by the additional $U(1)$ symmetry, so the shape of the rings does not modify the total energy of the solution.

We summarise the above results noticing that in the second model, with $U(1)\times U(1)$ symmetry, the first order equations \eqref{fovisible} and \eqref{fohidden} play a role which is similar to the case discussed before for the first model. Besides simplifying the search for solutions, they also obey the minimum energy condition which ensure their linear stability. In particular, we can also introduce a topological current and show that the corresponding topological charge is conserved and related to the flux of the magnetic field $B=-a^\prime/r $ described under the local $U(1)$ symmetry. Evidently, we can use another model related to the extra $U(1)$ symmetry. It can be, for instance, described by the Chern-Simons term, instead the Maxwell one that we used above. This case is harder, because the Chern-Simons dynamics makes the vortex charged electrically and requires the presence of cubic and quintic nonlinearities; see, e.g., Ref. \cite{Jackiw}. However, since in our model the extra $U(1)$ symmetry is included independently, this case can be implemented with no new obstacle, so we do not investigate this possibility in this work.

\section{Discussion}
\label{sec:end}

In this work we have studied vortices in a relativistic model with symmetry $U(1)\times G$, with $G$ being the $Z_2$ symmetry or another local $U(1)$ symmetry. In the two cases, we have developed first order framework which unveils the presence of stable vortex solutions, with the form of a central core surrounded by shells that are controlled by the parameters included in the extensions used to define the models. Although the enhancement of the local $U(1)$ symmetry to $U(1)\times Z_2$ or $U(1)\times U(1)$ is not new, here we innovate to describe the entrapment of vortices into geometrically constrained regions in the plane. In the case with the addition of the discrete $Z_2$ symmetry, we have examined two distinct possibilities, one with cubic nonlinearity, and the other with cubic and quintic nonlinearities, to help shed some light on the role played the by presence of nonlinearity. Anyway, nonlinearity related to the additional symmetry is mandatory to give rise to the localized structure to entrap the vortex. 

Despite the intrinsic differences between the models with $U(1)\times Z_2$ and $U(1)\times U(1)$ symmetries, the internal modification of the vortex is somehow similar, since it changes from a single central core to the form of a multilayered structure. We notice, however, that the total energy of the field configurations engenders distinct behavior: in the first model, it depends on $\alpha$, a parameter that appears connected with the additional discrete $Z_2$ symmetry; however, in the second model with additional $U(1)$ symmetry, it does not depend on $q$, the parameter that appear linked to the additional local gauge symmetry.

We believe that these novel structures are of current interest to applications dealing with planar magnetic core and shells structures, in particular with magnetic vortices, since they are similar to the bimagnetic core and shell nanometric structures that appear in space \cite{CS}, that can be more accurate to the construction of the new generation of devices with applications in magnetic resonance imaging and magnetic data recording, to quote just two possibilities. Since we are working in $(2,1)$ spacetime dimensions, one could consider the possibility to change the Maxwell term to the Chern-Simons one, to see if it is possible to modify the standard Chern-Simons vertex configuration \cite{Jackiw} into a multilayered structure of the form presented in this work. This is harder because the Chern-Simons vortex is charged electrically, engendering a constraint that must be solved appropriately to generate acceptable solution. However, a Chern-Simons model was investigated recently in \cite{V1} and the results motivate further investigation on this issue.

Another possibility of current interest concerns the extension of the present work to the case of magnetic monopoles. This will certainly need the non Abelian $SU(2)$ symmetry to control the gauge fields in the three-dimensional space. The problem here is much harder than the case of vortices examined above, but the recent investigations in which the magnetic monopole may change its standard form to the small or hollow \cite{SH} or the bimagnetic \cite{BM} shape motivate further investigations, aimed to transform the standard monopole into a multilayered structure. In the case of vortices, we can also think of the basic $U(1)$ symmetry to describe the visible sector, with the additional symmetry describing the hidden sector. This issue was investigated recently in \cite{Schapo,ahep} with different couplings, motivated to study possible mechanisms to describe interaction between the baryonic and dark matter within the low-energy frontiers of particle physics \cite{LEF}.

Other interesting possibilities concern the study of the nonlinear Schr\"odinger and Gross-Pitaevskii equations. It is a known fact that under specific conditions, they may support vortices of several distinct profiles \cite{NS,GP}; see, for instance, Refs. \cite{RVprl,A,B,C,PRL,PRLvortex,PRE} for some specific investigations on vortices guided by the nonlinear Schr\"odiger or the Gross-Pitaevskii equations. In \cite{RVprl}, for instance, the author investigated light propagation along the $z$ axis in a bulk medium with defocusing cubic nonlinearity and transverse modulation of the refractive index, capable of generating stable ringlike vortex configurations; see also Refs. \cite{A,B,C} for other related studies. The investigation described the presence of stable ringlike vortex configurations trapped by a potential that simulates a Bessel lattice, modulating the transverse profile of the refractive index that leads to the ringlike structures. We notice that there is a single equation, but there is a trapping potential which is controlled externally by the Bessel lattice. Differently, in \cite{PRE} the authors considered a binary Bose-Einstein condensate with tunable intercomponent interaction that is modulated periodically, with frequency that is resonant or nonresonant with the frequency of the harmonic trapping potential. The many possibilities gave rise to ringlike and a variety of exotic patterns. In this case, there are two equations, with tunable intercomponent interaction. In the models proposed in the present work, the ringlike structures are modulated by changing the magnetic permeability of the medium. Although this is different from the approaches described in \cite{RVprl} and \cite{PRE}, we think that the idea described in the present work can be extended to the nonlinear Schr\"odinger and Gross-Pitaevsky equations, bringing novelties to both the optical vortices and their close relatives, vortices in Bose-Einstein condensates. Some of the above issues are now under consideration, and we believe that the present research will foster further studies on the subject. 

\acknowledgments{This research is suported in part by Conselho Nacional de Desenvolvimento Cient\'\i fico e Tecnol\'ogico (Grants 306614/2014-6, 404913/2018-0, 130923/2018-4, 155551/2018-3, and 306504/2018-9) and by Paraiba State Research Foundation (Grants 0003/2019 and 0015/2019).}


\begin{thebibliography}{99}
\bibitem{Ab}A. A. Abrikosov, Zh. Eksp. Teor. Fiz. 32, 1442 (1957); Sov. Phys. JETP 5, 1174 (1957).
\bibitem{GL}V. L. Ginzburg and L. D. Landau, Zh. Eksp. Teor. Fiz. 20, 1064 (1950).
\bibitem{NO}H. B. Nielsen and P. Olesen, Nucl. Phys. B 61, 45 (1973).
\bibitem{Sci1}T. Shinjo, T. Okuno, R. Hassdorf, K. Shigeto, and T. Ono, Science 289, 930 (2000).
\bibitem{Sci2}A. Wachowiak, J. Wiebe, M. Bode, O. Pietzsch, M. Morgenstern, and R. Wiesendanger, Science 289, 577 (2002).
\bibitem{spinor}Y. Kawaguchi and M. Ueda, Phys Rep. 520, 253 (2012).
\bibitem{Sci3}L. V. Levitin, R. G. Bennett, A. Casey, B. Cowan, J. Saunders, D. Drung, Th. Schurig, and J. M. Parpia, Science 340, 841 (2013)
\bibitem{prb}J. J. Wiman and J. A. Sauls, Phys. Rev. B 92, 144515 (2015).
\bibitem{vb1}H. Wioland, F. G. Woodhouse, J. Dunkel, J. O. Kessler, and R. E. Goldstein, Phys. Rev. Lett. 110, 268102 (2013).
\bibitem{vb2}H. Wioland, F. G. Woodhouse, J. Dunkel, and R. E. Goldstein, Nat. Phys. 12, 341 (2016).
\bibitem{Bog}U. K. Rössler, A. N. Bogdanov, and C. Pfleiderer,  Nature 442, 797 (2006).
\bibitem{Nature}O. Janson, I. Rousochatzakis, A. A. Tsirlin, M. Belesi, A. A. Leonov, U. K. Rößler, J. van den Brink, and  H. Rosner,
Nat. Commun. 5, 5376 (2014).
\bibitem{NRM}A. Fert, N. Reyren and V. Cros, Nat. Rev. Mater. 2, 17031 (2017).
\bibitem{BLM}D. Bazeia, M. A. Liao, and M. A Marques, arXiv:1908.01085.
\bibitem{N1}E. Witten, Nucl. Phys. B 249, 557 (1985).
\bibitem{N2}M. Shifman, Phys. Rev. D 87, 025025 (2013).
\bibitem{N3}D. Bazeia, M.A. Marques, and R. Menezes, Phys Lett. B 780, 485 (2018).
\bibitem{N}Y. Zhou and M. Ezawa, Nat. Commun. 5, 4652 (2014).
\bibitem{S}W. Jiang, P. Upadhyaya, W. Zhang, G. Yu, M. B. Jungeisch, F. Y. Fradin, J. E. Pearson, Y. Tserkovnyak, K. L. Wang, O. Heinonen, S. G. E. te Velthuis, and A. Homann, Science 349, 283 (2015).
\bibitem{CP19}A. F. Schäffer, L. Rózsa, J. Berakdar, E. Y. Vedmedenko, and R. Wiesendanger, Communications Physics 2, 72 (2019).
\bibitem{D}I. Dzyaloshinsky, J. Phys. Chem. Solids 4, 241 (1958).
\bibitem{M}T. Moriya, Phys. Rev. 120, 91 (1960).
\bibitem{Sky}T. H. R. Skyrme, Nucl. Phys. 31, 556 (1962).
\bibitem{V1}D. Bazeia, M. A. Marques, and D. Melnikov, Phys. Lett. B 785, 454 (2018).
\bibitem{PRL} D. Bazeia, J. Menezes and R. Menezes, Phys. Rev. Lett. 91, 241601 (2003).
\bibitem{RVprl}Y. V. Kartashov, V. A. Vysloukh, and L. Torner, Phys. Rev. Lett. 94, 043902 (2005).
\bibitem{ahep} D. Bazeia, L. Losano, M. A. Marques, and R. Menezes, Adv. High Energy Phys. 2019, 3187289 (2019).
\bibitem{CS}A. L\'opez-Ortega, M. Estrader, G. Salazar-Alvarez, A. G. Roca, and J. Nogu\'es, Phys. Rep. 553, 1 (2015).
\bibitem{Jackiw}R. Jackiw and E. Weinberg, Phys. Rev. Lett. 64, 2234 (1990).
\bibitem{SH}D. Bazeia, M.A. Marques, and Gonzalo J. Olmo, Phys. Rev. D 98, 025017 (2018).
\bibitem{BM}D. Bazeia, M. A. Marques, and R. Menezes, Phys. Rev. D 98, 065003 (2018).
\bibitem{Schapo}P. Arias, E. Ireson, C. Nu\ nes, and F. Schaposnik, J. High Energy Phys. 1502, 156 (2015).
\bibitem{LEF}J. Jaeckel and A. Ringwald, Ann. Rev. Nucl. Part. Sci. 60, 405 (2010).
\bibitem{NS}A. S. Desyatnikov, Y. S. Kivshar, and L. Torner, {\it Optical vortex and vortex solitons.} Progress in  Optics, Vol. 47, Chapter 5 (North-Holland, 2005).
\bibitem{GP}C. J. Pethick and H. Smith, {\it Bose–Einstein Condensation in Dilute Gases.} (Cambridge University Press, 2008).
\bibitem{A}V. Garc\'es-Ch\'avez, D. McGloin, H. Melville, W. Sibbett, and K. Dholakia, Nature 419, 145 (2002).
\bibitem{B}Y. V. Kartashov, V. A. Vysloukh, and L. Torner, Phys. Rev. Lett. 93, 093904 (2004).
\bibitem{C}Y. V. Kartashov, A. Ferrando, A. A. Egorov, and L. Torner, Phys. Rev. Lett. 95, 123902 (2005).
\bibitem{PRLvortex}M. R. Matthews, B. P. Anderson, P. C. Haljan, D. S. Hall, C. E. Wieman, and E. A. Cornell
Phys. Rev. Lett. 83, 2498 (1999).
\bibitem{PRL}G. Theocharis, D. J. Frantzeskakis, P. G. Kevrekidis, B. A. Malomed, and Y. S. Kivshar, Phys. Rev. Lett. 90, 120403 (2003).
\bibitem{PRE}Z.-M. He, L. Wen, Y.-J. Wang, G. P. Chen, R.-B. Tan, C.-Q. Dai, and X.-F. Zhang, Phys. Rev. E 99, 062216 (2019).
\end{thebibliography}
\end{document}